\documentclass[useAMS,usenatbib]{mn2e}
\usepackage{graphicx,amsmath,txfonts}

\title{Orbital alignment of circumbinary planets that form in misaligned circumbinary discs: the case of Kepler-413b}
\author[A. Pierens, R.P. Nelson]{A. Pierens $^{1,2}$, R.P. Nelson$^{3}$ \\
$^1$Universit\'e de Bordeaux, Observatoire Aquitain des Sciences de l'Univers,
    BP89 33271 Floirac Cedex, France \\
$^2$CNRS, Laboratoire d'Astrophysique de Bordeaux,
     BP89 33271 Floirac Cedex, France\\
 $^3$  Astronomy Unit, Queen Mary University of London, Mile End Road, London, E1 4NS, UK}
\date{Released 2012 Xxxxx XX}

\pagerange{\pageref{firstpage}--\pageref{lastpage}} \pubyear{2014}

\def\LaTeX{L\kern-.36em\raise.3ex\hbox{a}\kern-.15em
    T\kern-.1667em\lower.7ex\hbox{E}\kern-.125emX}
\begin{document}
\label{firstpage}
\maketitle
\begin{abstract}
 Although most of the circumbinary planets detected by the \emph{Kepler} spacecraft are on orbits that are closely aligned with the binary orbital plane, the systems Kepler-413 and Kepler-453 exhibit small misalignments of $\sim 2.5^\circ$. One possibility is that these planets formed in a circumbinary disc whose midplane was inclined relative to the binary orbital plane. Such a configuration is expected to lead to a warped and twisted disc, and our aim is to examine the inclination evolution of planets embedded in these discs. We employed 3D hydrodynamical simulations that examine the disc response to the presence of a modestly inclined binary with parameters that match the Kepler-413 system, as a function of disc parameters and binary inclinations. The discs all develop slowly varying warps, and generally display very small amounts of twist. Very slow solid body precession occurs because a large outer disc radius is adopted. Simulations of planets embedded in these discs resulted in the planet aligning with the binary orbit plane for disc masses close to the minimum mass solar nebular, such that nodal precession of the planet was controlled by the binary. For higher disc masses, the planet maintains near coplanarity with the local disc midplane. Our results suggest that circumbinary planets born in tilted circumbinary discs should align with the binary orbit plane as the disc ages and loses mass, even if the circumbinary disc remains misaligned from the binary orbit. This result has important implications for understanding the origins of the known circumbinary planets.

\end{abstract}

\section{Introduction}
To date, $11$ circumbinary planets orbiting both components of a close binary system have been discovered by the {\it Kepler} spacecraft.  With the exception of Kepler-413b (Kostov et al. 2015) and Kepler-453b (Welsh et al. 2015) whose orbital planes are misaligned by $\sim 2.5^\circ$ with respect to the binary orbital planes, most of these planets have orbits that are closely aligned to those of the central binaries, supporting the idea that these planets formed in a circumbinary disc that was close to being coplanar with the central binary. 

A number of the \emph{Kepler} circumbinary planets are on orbits that are close to the region of dynamical instability near the central binary (Holman \& Wiegert 1999). Two scenarios have been put forward to explain the presence of planets in these orbits. In-situ formation, involving the accretion of planetesimals at the observed locations of the planets, has been suggested. However, various studies have highlighted the difficulty of building circumbinary planets in this way, due to the planetesimals being excited onto highly eccentric orbits (Paardekooper et al. 2012; Meschiari 2012 a,b; Lines et al. 2014; Bromley \& Kenyon 2015), size-dependent pericentre alignment of planetesimals (Scholl et al. 2007)  or gravitational interactions with non-axisymmetric eccentric features in the disc (Marzari et al. 2008; Kley \& Nelson 2010) all contributing to mutual collisions being disruptive rather than accretionary. Studies involving pebble accretion have yet to be undertaken, but one might expect that the large velocity perturbations induced in the gas and solid component of the disc in the inner regions would render planet formation via pebble accretion an inefficient process there.

 An alternative, more plausible,  possibility is that these planets formed further out in the disc, in a more accretion-friendly environment, and migrated to their present location through Type I (Ward et al. 1997; Tanaka et al. 2002) and/or Type II migration (e.g. Lin \& Papaloizou 1986; Nelson et al. 2000).  Previous studies of the interaction of  planets with circumbinary discs (Nelson 2003; Pierens \& Nelson 2007, 2008a,b, 2013; Kley \& Haghighipour 2014, 2015) have indeed shown that migrating circumbinary planets naturally park at the edge of the central cavity formed by the binary. 
 
The fact that the circumbinary planets detected so far evolve on orbits that are coplanar with the binary is at first sight consistent with the general expectation that the angular momentum vectors of the circumbinary disc and binary should be primordially aligned because both originate from a single cloud. We note, however, that in the context of the {\it Kepler} mission, it is of course easier to detect planets whose orbits are aligned with the binary (Winn et al. 2015), since even a small misalignment can prevent the planet from transiting for long periods of time (precession does, however, lead to times when even a misaligned planet transits one or both stars). Hence, the lack of misaligned circumbinary planets may simply be due to selection bias. A primordial misalignment between the disc midplane and the binary orbit, that could in principle lead to the formation of a misaligned circumbinary planet, might arise due to turbulence the cloud (Padoan \& Nordlund 2002),  leading to infall of gas whose angular momentum vector varies with time (Bate et al. 2003). This would  be consistent with the observation of inclined circumbinary discs, such as around the pre-main-sequence star binary KH 15D (Chiang et al. 2004), or the binary protostar IRS 43 (Brinch et al. 2016). 

The misalignment between the circumbinary disc and orbit plane of the binary leads to disc warping (Larwood \& Papaloizou 1997), to a degree that depends strongly on the disc aspect ratio, $h$, and viscosity parameter, $\alpha$. In the case where  $h>\alpha$ and the warp propagation time is shorter than the differential precession timescale, the disc undergoes rigid precession with a small warp, with a precession frequency that tends to  zero if the disc is arbitrarily large (Larwood \& Papaloizou 1997). Stronger warps develop in the diffusive regime for warp propagation (i.e. when $\alpha>h$) and/or if the warp propagation timescale is longer than the differential precession timescale (Larwood et al 1996; Fragner \& Nelson 2010). 

Such a disc warp may  have a significant impact on the inclination evolution of embedded planets. Terquem (2013) has shown that secular interaction with a warped disc can make the planetary orbit precess, and  can even lead to very high inclinations when the warp is strong enough. For a disc that is inclined relative to a binary companion, the outcome of the secular interaction between the disc and an embedded planet depends mainly on the disc mass and the ability for the planet to open a gap in the disc. Coplanarity between the disc midplane and the planetary orbit can be maintained if the disc is massive enough (Xiang-Gruess \& Papaloizou 2014; Lubow \& Martin 2016),  whereas the planet tends to decouple from the disc in low-mass discs or when it opens a gap in the disc (Picogna \& Marzari 2015). In the context of circumbinary discs, Foucart \& Lai (2013) have shown that the back-reaction torque exerted by a warped disc on the binary tends to align the disc midplane with the binary. Because the alignment timescale increases with the binary semi-major axis,  they suggest that wide binaries with separations $\gtrsim 1$ AU can have misaligned circumbinary discs whereas close binaries should be aligned with their circumbinary disc. The implication is that, provided they formed in the late phase of the circumbinary disc, the circumbinary planets should evolve on orbits which are aligned with the binary, consistently with what is currently observed.

In this paper, we employ 3D hydrodynamical simulations to investigate whether or not, for the Kepler-413b  circumbinary planet, the small misalignment between the planetary and binary orbits could have resulted from migration in a circumbinary disc whose midplane is initially misaligned with respect to the binary orbit. The first issue that we address is the hydrodynamic response of the disc to the precessional torque exerted by the binary, as a function of disc parameters and binary inclination. We show that the disc tends towards a quasi-steady state displaying essentially no precession, and generally develops a small warp (i.e. one in which the rate of change of the disc inclination with radius is small), such that the outer disc remains in its original orbit plane, but the inner disc settles towards a plane that is modestly inclined with respect to the binary plane. We use a subset of these relaxed disc models as initial conditions for simulations that examine the orbital evolution of planets that are subject to gravitational interactions with the disc and binary. We find that coplanarity between the planet orbit and the disc midplane is maintained for high mass discs. As the disc mass is decreased, however, the planet decouples gravitationally from the disc and becomes dominated by the binary, which drives precession out of the disc midplane. Tidal interaction between the disk and planet leads to damping of the planet inclination, and the planet evolves towards coplanarity with the binary, as this provides a plane of symmetry for the system. Hence, we suggest that as an inclined circumbinary disc containing a planet on a short period evolves, such that its mass decreases due to accretion onto the central stars or perhaps through mass loss in a wind, a time will be reached when the planet will start to align with the binary. Given sufficient time, it will settle into the binary plane. Hence, even misaligned circumbinary discs may give rise to short period planets on orbits that are coplanar with the binary. 

This paper is organized as follows. In Section 2 we describe the hydrodynamical model and the initial conditions that are used in the simulations. In Section 3 we present the results of simulations of circumbinary discs that are modestly misaligned with respect to the binary orbit. The orbital evolution of embedded planets that are initially coplanar with the disc is then studied in Section 4. Finally, we discuss our results and draw our conclusions in Section 5.

\section{The hydrodynamic model}
\subsection{Numerical setup}
The simulations presented in this paper were performed using FARGO3D (Benitez-Lamblay \& Masset 2016). This code uses finite differences, and an advection scheme based on the monotonic transport algorithm (Van Leer 1977).  The FARGO algorithm (Masset 2000) is employed to avoid time step limitations due to the Keplerian velocity at the inner edge of the disc.
 
We adopt locally isothermal disc models, and solve the following equations for the conservation of mass and momentum in spherical coordinates $(r,\theta,\phi)$ with the origin of the frame located at the centre of mass of the binary,  
where $r$ is the radial distance from the origin, $\theta$ is the polar angle measured from the z-axis and $\phi$ is the 
azimuthal coordinate starting from the x-axis:
 \begin{equation}
 \frac{\partial \rho}{\partial t}+{\bf \nabla}\cdot(\rho {\bf v})=0,
 \end{equation}
 \begin{equation}
 \rho\left(\frac{\partial {\bf v}}{\partial t}+{\bf v}\cdot{\bf \nabla} {\bf v}\right)=-{\bf \nabla} P-\rho {\bf \nabla} \Phi +{\bf \nabla} \cdot {\bf \Pi},
 \end{equation}
where $\rho$ is the density, ${\bf v}$ the velocity, P the pressure, ${\bf \Pi}$ the viscous stress tensor and $\Phi$ is the gravitational potential, which can be written as:
 \begin{equation}
 \Phi=-\frac{GM_1}{|\mathbf{r}-\mathbf{r_1}|}-\frac{GM_2}{|\mathbf{r}-\mathbf{r_2}|},
 \end{equation}
where $M_1$, $M_2$ are the masses, and ${\bf r_1}$, ${\bf r_2}$  the radius vectors of the primary and secondary stars respectively. We do not include disc self-gravity in this work. Previous work (Mutter et al. 2017) for flat 2D discs showed that for disc masses $\lesssim 5$ MMSN (where MMSN refers to the minimum mass solar nebula model of Hayashi (1981)), the disc structure is only weakly affected by self-gravity. Here, our reference disc mass corresponds to $1$ MMSN, so development of the disc structures reported by Mutter et al.(2017) should not arise. Self-gravity, however, can lead to changes in the precession rates of discs, and this could in principle modify the disc structure. We leave exploration of this effect to future work.

When a planetary companion is included, it contributes to the gravitational potential through the expression \begin{equation}
 \Phi_p=-\frac{Gm_p}{(|\mathbf{r}-\mathbf{r_p}|^2+b^2)^{1/2}}
 \end{equation}
where $m_p$ is the planet mass, ${\bf r_p}$ is the planet radius vector and $b$ is the smoothing length which corresponds approximately to the cell diagonal size. 

Computational units are chosen such that  the total mass of the binary is $M_{\rm bin}=1$, the gravitational constant $G=1$, and the radius $r=1$ in the computational domain corresponds to the binary semi-major axis for the  Kepler-413 system ($a_{bin}\sim 0.1$ AU, see Table 1). When presenting simulation results, we use the binary orbital period $T_{bin}=2\pi\sqrt{a_{bin}^3/GM_{bin}}$ as the unit of time. The computational domain extends from $R_{in}=1.5a_{bin}$ to $R_{out}=80a_{bin}$. Here, employing a large disc model is important  because the disc precession frequency strongly depends on the value for $R_{out}$. For an effectively infinite disc that is inclined with respect to the binary orbital planet, we would expect  the disc to maintain a steady warped structure without any global precession, and which globally aligns to the binary orbit on a viscous time scale. Because of the large disc radial extent, we employ a logarithmic grid using $574$ radial grid cells. In the azimuthal direction the simulation domain extends from $0$ to $2\pi$ with $700$ uniformly spaced grid cells.  In the meridional direction, the simulation domain covers  $E[5+\gamma_{bin}/h_{max}]$ disc pressure scale heights above and below the disc midplane, where $E[x]$ denotes the nearest integer to the real $x$,  $\gamma_{bin}$ is the binary inclination and $h_{max}$ is the aspect ratio at the outer edge of the disc.

\subsection{Initial conditions}
In this work, we focus on locally isothermal discs for which the aspect ratio does not vary with time and is given by $h=h_0(r/r_0)^f$,  where $f$ is the disc flaring index and $r_0=1$. Most of the models have constant 
aspect ratio with $h_0=0.05$, but we also considered a flared disc model with $f=0.25$ and $h_0=0.01$. This particular model was designed in a such a way that a planet with mass equal to that of Kepler-413b (see Table 1) would be able to carve a gap in the disc when located at $a_p\sim 3.5 a_{bin}$.

The effective kinematic viscosity $\nu$ is modelled using the standard alpha prescription  $\nu=\alpha c_s H$ (Shakura \& Sunyaev 1973) where  $c_s$ 
the sound speed and $\alpha$ the viscosity parameter. In this work we explore how changing $\alpha$ modifies the disc structure, and adopt the values $\alpha=10^{-5}$, $4\times 10^{-3}$ and $10^{-1}$.

We consider circumbinary disc whose midplane start off in the equatorial plane of the computational grid, with a binary whose orbital plane is inclined.  The initial density and azimuthal velocity profiles satisfy hydrostatic equilibrium and are given by:
\begin{equation}
\rho(r,\theta)=f_{gap}\rho_0\left(\frac{r}{r_0}\right)^{-\xi}\left[\sin (\theta) \right]^{\frac{1}{h^2}+2f-(\xi+1)}
\end{equation}
and 
\begin{equation}
v_\phi(r,\theta)=\left[\frac{GM}{r}\left(1+(2f-(\xi+1))h^2\right)\right]^{1/2},
\end{equation}
\label{sec:initial}
 with $\rho_0$  the density at $r_0=1$ and with $\xi=1+p+f$, where $p=3/2$ and $\beta=1-2f$ are the power-law indexes for the surface density and temperature, respectively. $\rho_0$ is set to a value corresponding to a $1$ MMSN disc, namely containing $2\%$ of the mass of the binary within $30$ AU. $f_{gap}$ is a gap-function used to initiated the disc with an inner cavity (created by the binary), and is given by:
\begin{equation}
f_{gap}=\left(1+\exp\left[-\frac{r-r_{gap}}{0.1r_{gap}}\right]\right)^{-1},
\end{equation}
 where $r_{gap}=2.5a_{bin}$ is the analytically estimated gap size (Artymowicz \& Lubow 1994).
The binary semi-major axis is set to $a_{bin}=1.0$ while the binary eccentricity $e_{bin}$ and mass ratio $q_{bin}$ are set to match those of Kepler-413b (see Table 1).  Regarding the binary inclination $\gamma_{bin}$, we explored four cases with values  $0^\circ \le \gamma_{bin} \le 8^\circ$. In each case, the ascending node of the binary is located on the $x$-axis. The gravitational back-reaction from the disc onto the binary is not included, so the binary orbit remains fixed. Including the effect of the torque exerted by the disc onto the binary would be expected to lead to a decrease of the binary semi-major axis and evolution of its eccentricity (Pierens \& Nelson 2007; Fleming \& Quinn 2017). On time scales that depend on the disc mass, alignment of the orbital angular momentum of the binary with that of the disc would also arise (e.g. Foucart et Lai 2013). In total, we considered a set of 7 different disc models that are characterized by their values for $\alpha$, $\gamma_{bin}$, $h_0$, $f$. Details of parameters for each run can be found in Table $2$.

\begin{table}
\caption{Binary and planet parameters (from Kostov et al. 2015).}              
\label{table1}      
\centering                                      
\begin{tabular}{c c}          
\hline\hline                        
 & Kepler 413    \\ 
\hline 
$M_1$ $(M_\odot)$ & 0.82\\
$M_2$ $(M_\odot)$ & 0.54\\
$q_{bin}=M_2/M_1$ & 0.66\\
$q_p$ & $1.6\times 10^{-4}$\\                           
$a_{bin}$ (AU) & $0.10$  \\
$a_p$ (AU) & $0.35$ \\     
$e_{bin}$ & $0.04$  \\
$e_p$ & $0.12$ \\
$i_p$ $(^\circ)$ &  2.5\\
\hline                                             
\end{tabular}
\end{table}

\begin{table*}
\caption{Binary and planet parameters considered in our runs. }              
\label{table2}      
\centering    
\resizebox{0.7\textwidth}{!}{                                    
\begin{tabular}{ccccccc}          
\hline\hline                        
  Run label & $\alpha$ & $\gamma_{bin}$ & $h_0$ & f & Resolution   &   Meridional extent\\ 
  &  & &  &  &  ($N_r \times N_\phi \times N_\theta)$  &   \\ 
\hline                                   
 $1$ & $4\times10^{-3}$ & $5^\circ$ & $0.05$ & $0$ & $574\times700\times70$ &  $40^\circ$\\ 
 $2$ & $4\times10^{-3}$ & $2.5^\circ$ & $0.05$ & $0$ & $574\times700\times60$ &  $34^\circ$\\
 $3$ & $4\times10^{-3}$ & $8^\circ$ & $0.05$ & $0$ & $574\times700\times80$ &  $46^\circ$\\
  $4$ & $4\times10^{-3}$ & $0^\circ$ & $0.05$ & $0$ & $574\times700\times60$ &  $34^\circ$\\
 $5$ & $10^{-4}$ & $2.5^\circ$ & $0.05$ & $0$ & $574\times700\times60$ &  $34^\circ$\\
  $6$ & $10^{-1}$ & $2.5^\circ$ & $0.05$ & $0$ & $574\times700\times60$ &  $34^\circ$\\
 $7$ & $4\times10^{-3}$ & $5^\circ$ & $0.01$ & $0.25$ & $574\times700\times140$ &  $40^\circ$ \\  
\hline                                             
\end{tabular}
}
\end{table*}

\subsection{Boundary conditions}
We employ a closed boundary condition at the outer edge of the disc, and a viscous outflow condition (Pierens \& Nelson 2008) at the inner edge, for which  the radial velocity in the ghost zones is set to  $v_r=\beta v_r(R_{in})$, where $v_r(R_{in})=-3\nu/2R_{in}$ is the 
gas drift velocity due to viscous evolution and $\beta$ is a free parameter which is set to $\beta=5$.  In a recent study, Mutter et al. (2016) showed that the radial structure of a circumbinary disc (e.g. surface density profile, tidally truncated cavity size) is robust regarding the choice of the inner boundary condition only when the inner edge of the computational domain is small enough for the secondary star to be completely embedded within the computational domain. The computational expense incurred by 3D hydrodynamic simulations prevents us from  adopting such a small value for the inner radius because of the time step constraint that it would impose, so the structure of the tidally truncated cavity may be impacted by our  choice for the inner boundary condition (Kley \& Haghighipour 2014; Mutter et al. 2016). This issue is clearly of importance if trying to predict the final stopping location of a migrating planet, but in this study we are focussing on how the planet inclination evolution varies with disk mass. The precise form of the radial density distribution should not be so important when addressing that issue. Outflow boundary conditions are used at the meridional boundaries, for which all quantities in the ghost zones have the same values as in the first active zones, except the meridional velocity whose value is set to $0$ if it is directed towards the disc midplane to prevent inflow of material.

\section{Warp evolution in circumbinary discs}
\begin{figure}
\centering
\includegraphics[width=0.48\textwidth]{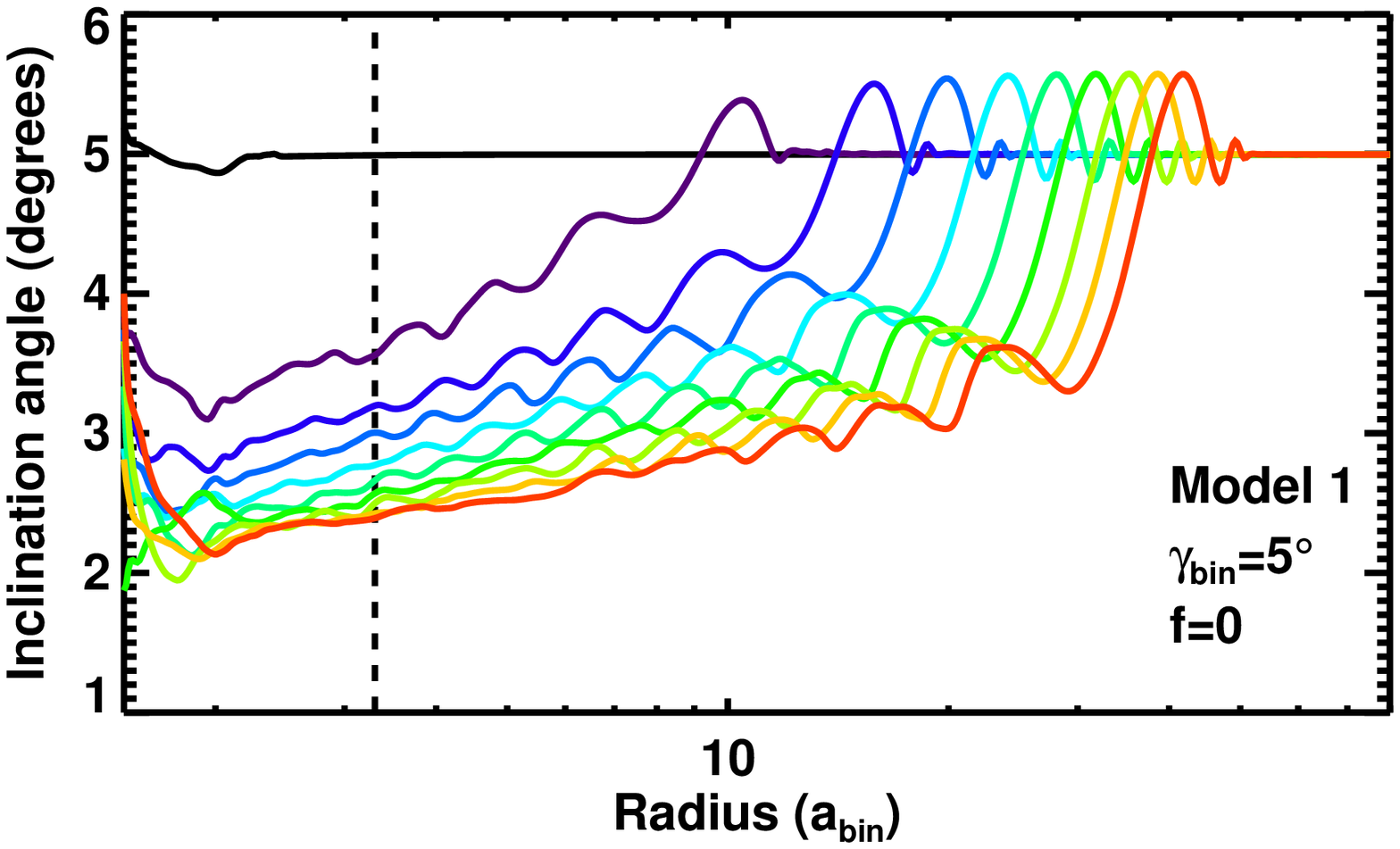}
\includegraphics[width=0.48\textwidth]{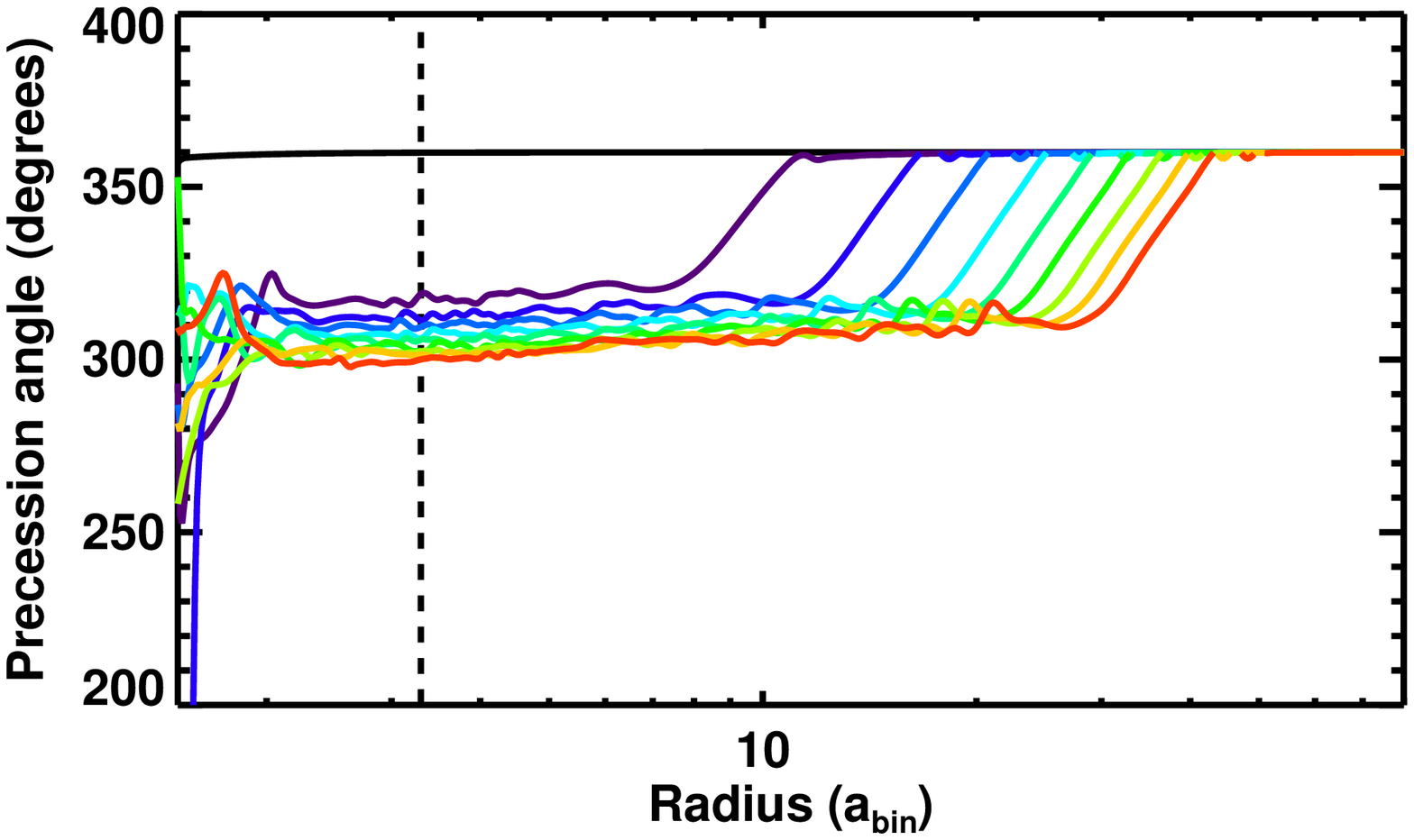}
\caption{{\it Upper panel}: Disc inclination angle $i_D$ as a function of radius at times $t=1, 100, 200, ...$ for Model $1$ 
which has constant aspect ratio with $h=0.05$,  viscosity parameter $\alpha=4\times 10^{-3}$, and binary inclination $\gamma_{bin}=5^\circ$.  Evolution proceeds from left to right on this figure. The vertical dashed line shows the current location of Kepler-413b. {\it Bottom panel:} Disc precession angle $\beta_D$ as a function of radius at the same times.}
\label{fig:model1}
\end{figure}

We now describe the warp evolution for each of the misaligned circumbinary disc models we considered.  In the following, the inclination angle, $i_d$, between the angular momentum vectors of disc annuli located at radius $r$ and the binary is defined as:
\begin{equation}
\cos(i_d(r))=\frac{{\bf L_D}\cdot {\bf L_B}}{|{\bf L_B}||{\bf L_D}|}
\end{equation}
where ${\bf L_B}$ is the binary orbital angular momentum vector and
 ${\bf L_D}$ is the angular momentum of disc annuli located at radius $r$:
 \begin{equation}
 {\bf L_D}=\int_0^{2\pi}\int_{\pi/2-\Delta \theta}^{\pi/2+\Delta \theta}{\bf l} r^2\sin(\theta)drd\theta d\phi,
 \end{equation}
 where $dr$ is the radial grid spacing and $\bf l=\rho ({\bf r}\times {\bf v})$ the angular momentum density. Similarly, we define the precession angle 
 of the disc angular momentum vector around ${\bf L}_B$ as:
 \begin{equation}
 \cos(\beta_d(r))=\frac{{\bf L_D}\times {\bf L_B}}{|{\bf L_B}\times {\bf L_D}|}\cdot {\bf u},
 \end{equation}
 where ${\bf u}$ is a reference unit vector in the orbital plane of the binary, defined to be ${\bf u}={\bf e}_x$. 
 
A disc annulus located at $r$ undergoes precession at a frequency given approximately by the 
free particle precession rate due to the torque exerted by the binary (e.g. Nixon et al. 2013):
\begin{equation}
\Omega_{prec}(r)=-\frac{3\eta}{4} \frac{a_b^2}{r^2}\Omega(r)\cos(i_D(r))
\label{eq:omegap}
\end{equation}
where $\eta=M_1M_2/M_{bin}^2$ and $\Omega$ is the Keplerian frequency. Differential precession of neighbouring annuli induces a twist that causes their midplanes to become misaligned.  This generates pressure-induced horizontal shear motions which, in the case with $h>\alpha$,  are responsible for bending waves propagating  across the disc at speed $c_s/2$. In the case $h<\alpha$, however, the horizontal shear resulting from misalignment of disc midplanes is damped locally and a warp that is built up in the disc will evolve diffusively in that case.

\subsection{Evolution for a fiducial run: Model 1}
\label{sec:model1}
Figure \ref{fig:model1} displays the evolution of the disc inclination and precession angles, $i_D$ and $\beta_D$,  as a function of radius for Model 1 at  equally spaced times $t=1$, 100, 200, ...  up to the final run time $t=10^3$ $T_{bin}$.  Warp evolution takes place in the wave-like regime since  $\alpha<h$. It is clear that the system evolves towards a quasi-steady state with very small warp (i.e. changes in inclination occur smoothly over a large range of radii), due to bending waves communicating information about the local warp structure efficiently across the disc. At $t=10^3 \; T_{bin}$, we see that an equilibrium has been reached at $r \sim 3.5 a_{bin}$, the current location of Kepler-413b,  while the outer parts of the disc are still evolving because of the very long differential precession times there. It is interesting to notice that at this location, the inclination angle at equilibrium $i_D\sim 2.5^\circ$ is very close to the orbital misalignment of Kepler-413b relative to the binary orbital plane (see Table 1). \\
Here, we note that the fact that the disc is distorted up to 
$\sim 50\; a_{bin}$ at $t=10^3 \; T_{bin}$ is not  a consequence of the bending waves propagating from the inner disc but  rather simply results from disc precession.  Given that the warp propagates at speed $c_s/2$, we would indeed expect the outer regions of the disc located at $\sim 50 \;a_{bin}$ to respond to the changing inclination of the inner disc, due to transient bending waves, after  a time longer than $\sim 10^3 \, T_{bin}$.   In fact, our results  suggest that  only a small amount of precession is required for the disc to respond by setting up internal stresses to ensure uniform precession. The consequence is that  on a timescale corresponding to the differential precession time in the outer disc, we would expect a steady-state warped structure to develop across the disc, where the binary torques that try to twist up and warp the disc are counter balanced by internal stresses involving pressure and viscosity. On even longer timescales, we would expect the disc and binary orbit planes to align. For a fixed binary orbit, as considered here, where realignment occurs because the inner disc brings the outer disc into alignment with the binary orbit plane, this is likely to occur on the global viscous evolution time of the disc once the steady-state warped structure has been established. 
 
 In the wave-like regime for warp propagation, we note that a significant warp may eventually be established in the innermost regions where the differential precession induced by the binary is stronger, provided that the precession  timescale there $t_{prec}\sim  \Omega_{prec}^{-1}$ is shorter than the warp propagation timescale $t_w=2r/c_s$. Using Eq. \ref{eq:omegap}, we find $t_w<t_{prec}$ for $r > 3.6\;a_{bin}$. As discussed later in the paper (see Sect. \ref{sec:varyibin}), 
this corresponds roughly to the tidal truncation radius  such that we  would expect the whole disc to remain smooth and only be modestly warped, consistently with the simulations.  Also, we see in the bottom panel of Fig. \ref{fig:model1} that the system evolves towards a \emph{very} slowly precessing configuration with almost no twist (i.e. $\beta_D$ is $\sim$ constant), which corresponds to what would be expected for an infinite disc with $t_w<t_{prec}$ everywhere. In fact, when the latter condition is fullfilled, a disc with outer radius $R_{out}$ should precess as a rigid body with precession rate given by (Larwood \& Papaloizou 1997):
 \begin{equation}
 \omega_{prec}=-\frac{3}{4}G\eta a_{bin}^2\cos(\gamma_{bin})\frac{\int_{R_{in}}^{R_{out}}\Sigma r^{-2}dr}{\int_{R_{in}}^{R_{out}}\Omega\Sigma r^{3}dr},
 \end{equation}
 which we can write, for a surface density profile $\Sigma \propto r^{-3/2}$:
  \begin{equation}
  \omega_{prec} \simeq -\frac{3}{10} \eta \cos{\gamma_{bin}} \left(\frac{a_{bin}}{R_{out}}\right) \left(\frac{a_{bin}}{R_{in}}\right)^{5/2} \Omega_{bin}
 \end{equation}  
  For a finite disc with $R_{out}=80a_{bin}$, the previous relation gives  $\omega_{prec}\sim 10^{-4}\Omega_{bin}$. The results of the simulations, showing a very small precession rate,  are 
 clearly in very good agreement with such an expectation. These numbers, however, also demonstrate the extreme difficulty of simulating circumbinary systems with realistic outer disc radii over run times that allow the long-term, global, quasi-steady evolution to be examined. Nonetheless, as our simulations demonstrate, the more rapid evolution of the inner disc regions towards a quasi-steady state allows analysis of the dynamics there to be captured at reasonable computational cost.
 
\subsection{Dependence on binary inclination: Models 2, 3, 4}
\label{sec:varyibin}
\begin{figure}
\centering
\includegraphics[width=0.48\textwidth]{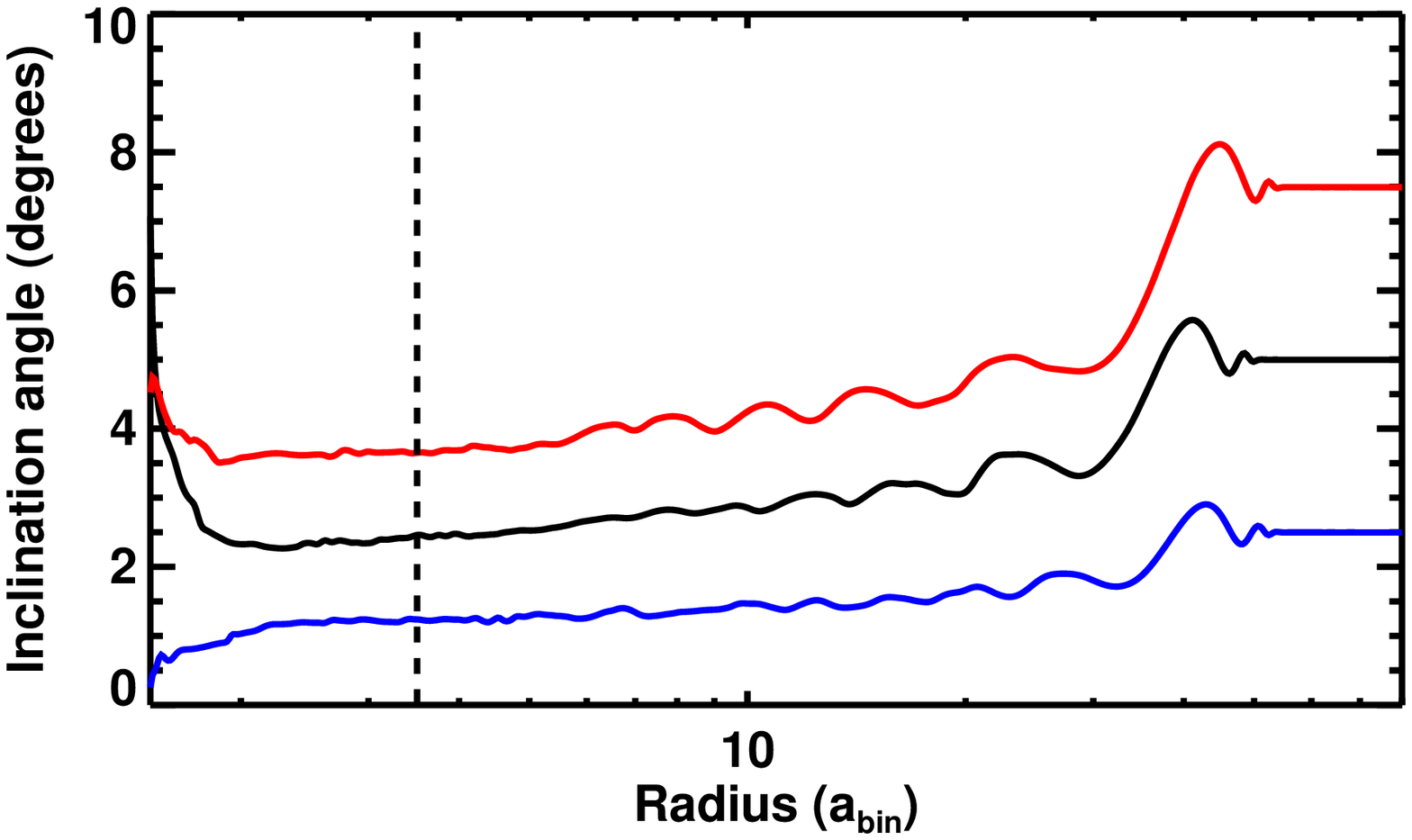}
\includegraphics[width=0.48\textwidth]{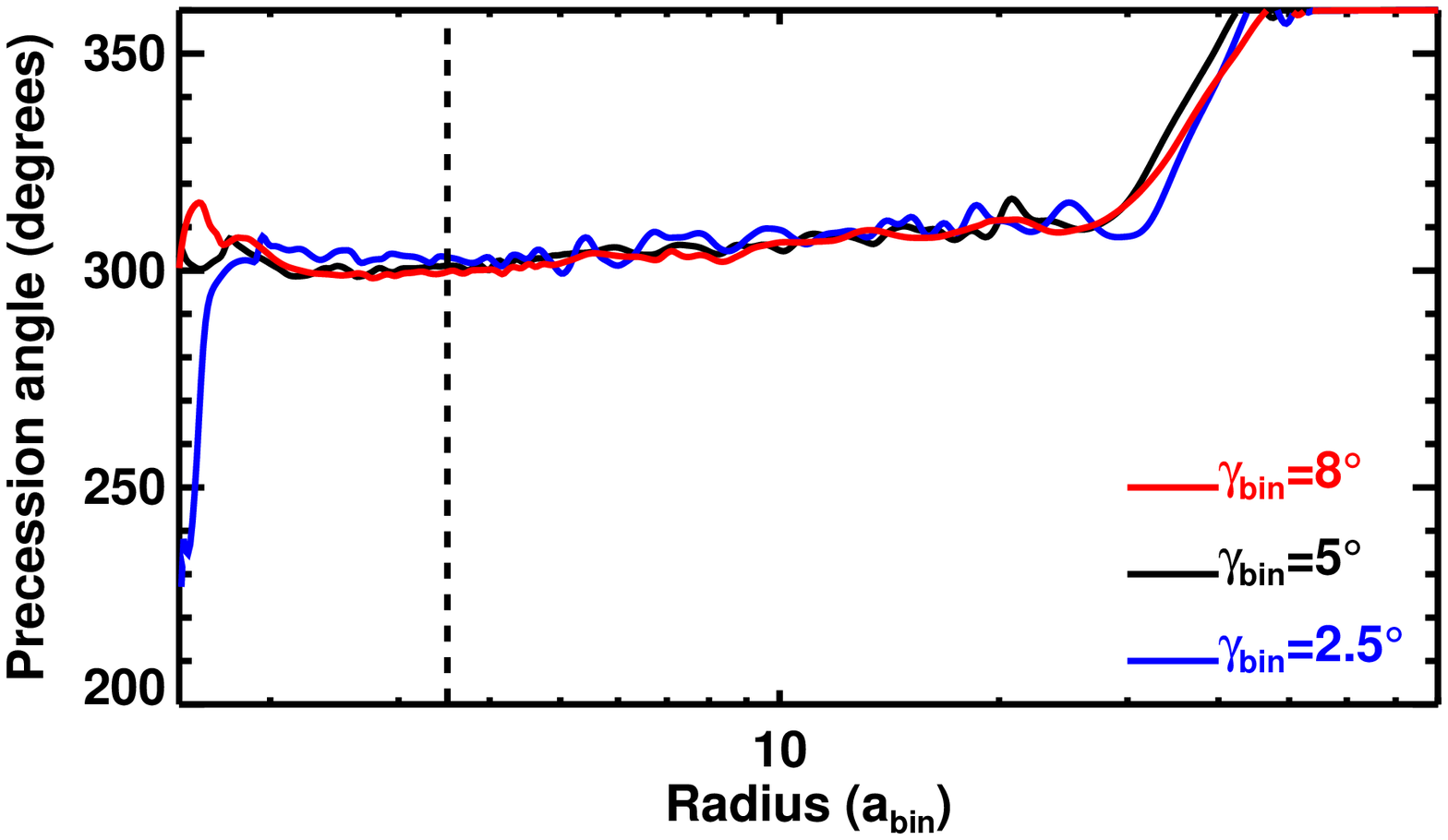}
\caption{{\it Upper panel:} Disc inclination angle $i_D$ as a function of radius at $t=1000$ $T_{bin}$, namely once a steady state is reached at the current location of Kepler-413b (vertical dashed line), for Models $1$, $2$, $3$ which have $\gamma_{bin}=5^\circ, 2.5^\circ, 8^\circ$ respectively. {\it Lower panel:} Disc precession angle $\beta_D$ as a function of 
radius at quasi-steady state for the same models. }
\label{fig:model23}
\end{figure}

We now examine how changing the binary inclination, $\gamma_{bin}$, influences the evolution. Although not shown here, we find that for Models $2$ and $3$, that have $\gamma_{bin}=2.5^\circ$ and $8^\circ$, respectively, the inclination and precession angles again reach constant values in the innermost parts of the disc at $t\sim 1000$ $T_{bin}$.  For these models, the inclination and precession angles at equilibrium are plotted as a function of radius  in Fig. \ref{fig:model23}. The weak dependence of  the differential precession rate on the binary inclination (see Eq. \ref{eq:omegap}) implies that for the range of values that we consider ($2.5\le \gamma_{bin} \le 8^\circ$), the relaxed structure of the disc should not vary too much from one model to another,  and this is indeed what we observe.  Fig. \ref{fig:model23} reveals that (i)  the final value for the precession angle does not depend on $\gamma_{bin}$, and (ii) a very small warp is established with $i_D\sim \gamma_{bin}/2$ at  $r=3.5 a_{bin}$ which is about the tidal truncation radius.

\begin{figure}
\centering
\includegraphics[width=0.48\textwidth]{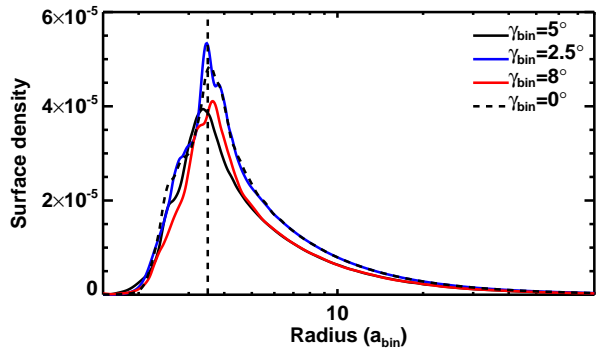}
\caption{Surface density profile at $t=1000$ $T_{bin}$, namely once a steady state is reached at the current location of Kepler-413b (vertical dashed line), for Models $1$, $2$, $3$, $4$ which have $\gamma_{bin}=5^\circ, 2.5^\circ, 8^\circ, 0^\circ$ respectively.}
\label{fig:model23profile}
\end{figure}

The azimuthally-averaged surface density profiles at $t=1000$ $T_{bin}$ are displayed in Fig. \ref{fig:model23profile}, and indicate that the size of the tidally cleared cavity and location of the maximum surface density are similar for all models.
For the disc parameters that we adopted,  it is somewhat encouraging that the location of the density peak in each model almost coincides with the current location of Kepler-413b, since it is  expected that the inward migration of a protoplanetary core will be stopped at the tidally truncated cavity  due to the action of corotation torques (Pierens \& Nelson 2007). 

\subsection{Dependence on $\alpha$ viscosity: Models 5, 6}
\begin{figure*}
\centering
\includegraphics[width=0.48\textwidth]{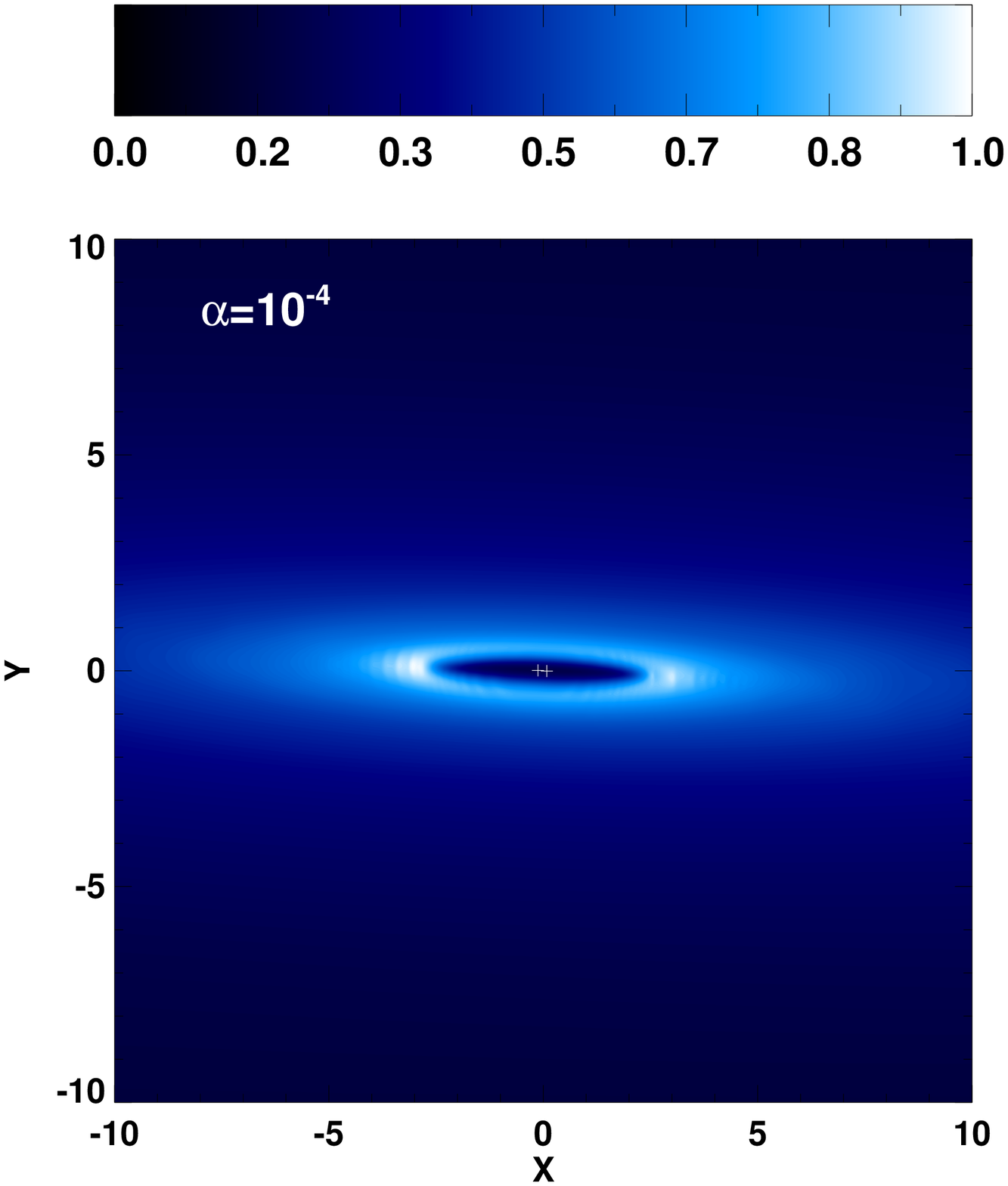}
\includegraphics[width=0.48\textwidth]{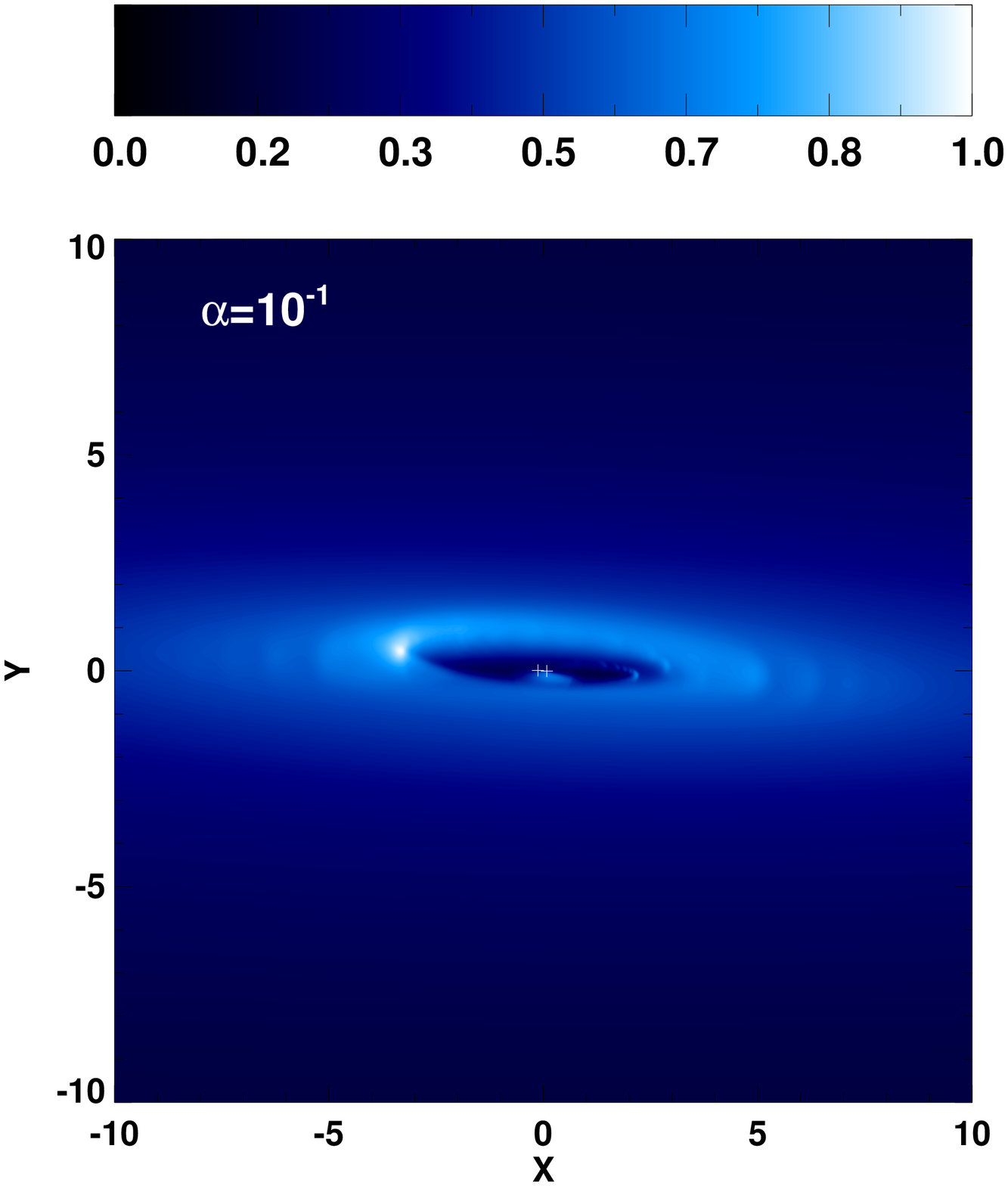}
\includegraphics[width=0.48\textwidth]{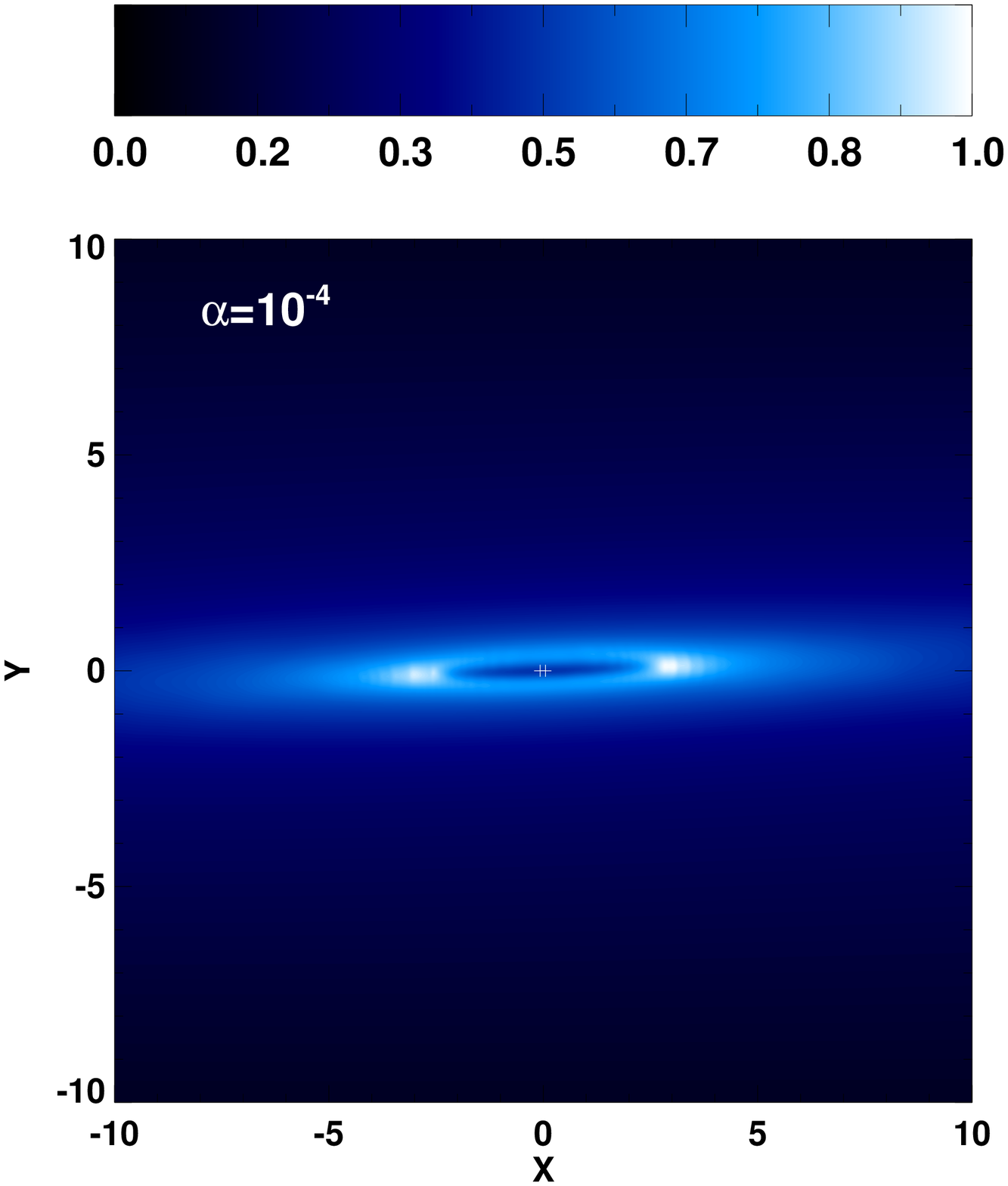}
\includegraphics[width=0.48\textwidth]{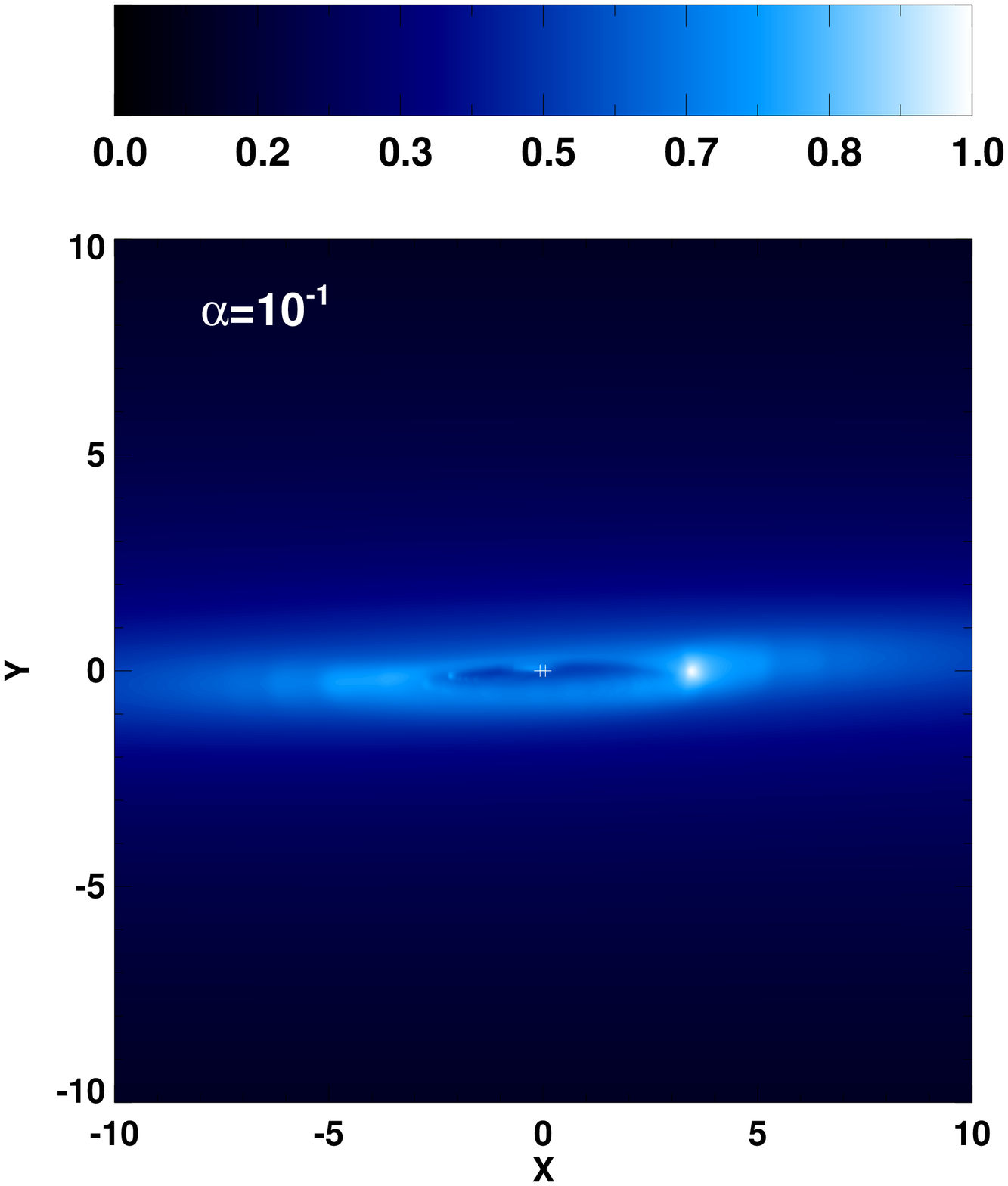}
\caption{ {\it Upper panel}: projected disc surface density along the line of sight characterized by the angles $(\theta,\phi)=(10^\circ, 80^\circ)$ for Model $5$ which has $\alpha=10^{-4}$ (left) and Model $6$ which has $\alpha=10^{-1}$ (right),  at $t=700$ $T_{bin}$. {\it Lower panel:} same but for a line of sight which is characterized by the angles $(\theta,\phi)=(185^\circ, 85^\circ)$. }
\label{fig:warp3d}
\end{figure*}

\begin{figure*}
\centering
\includegraphics[width=0.48\textwidth]{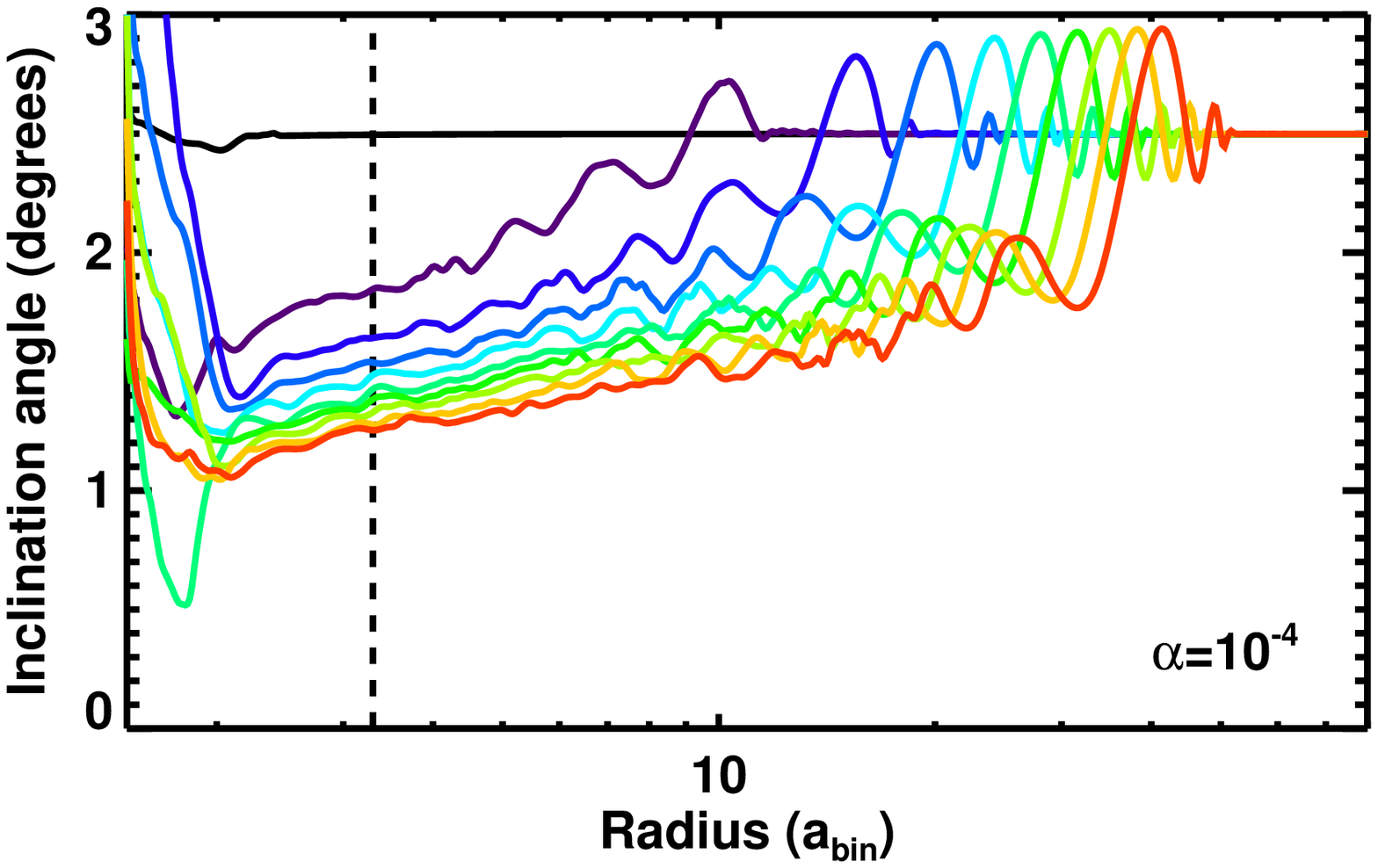}
\includegraphics[width=0.48\textwidth]{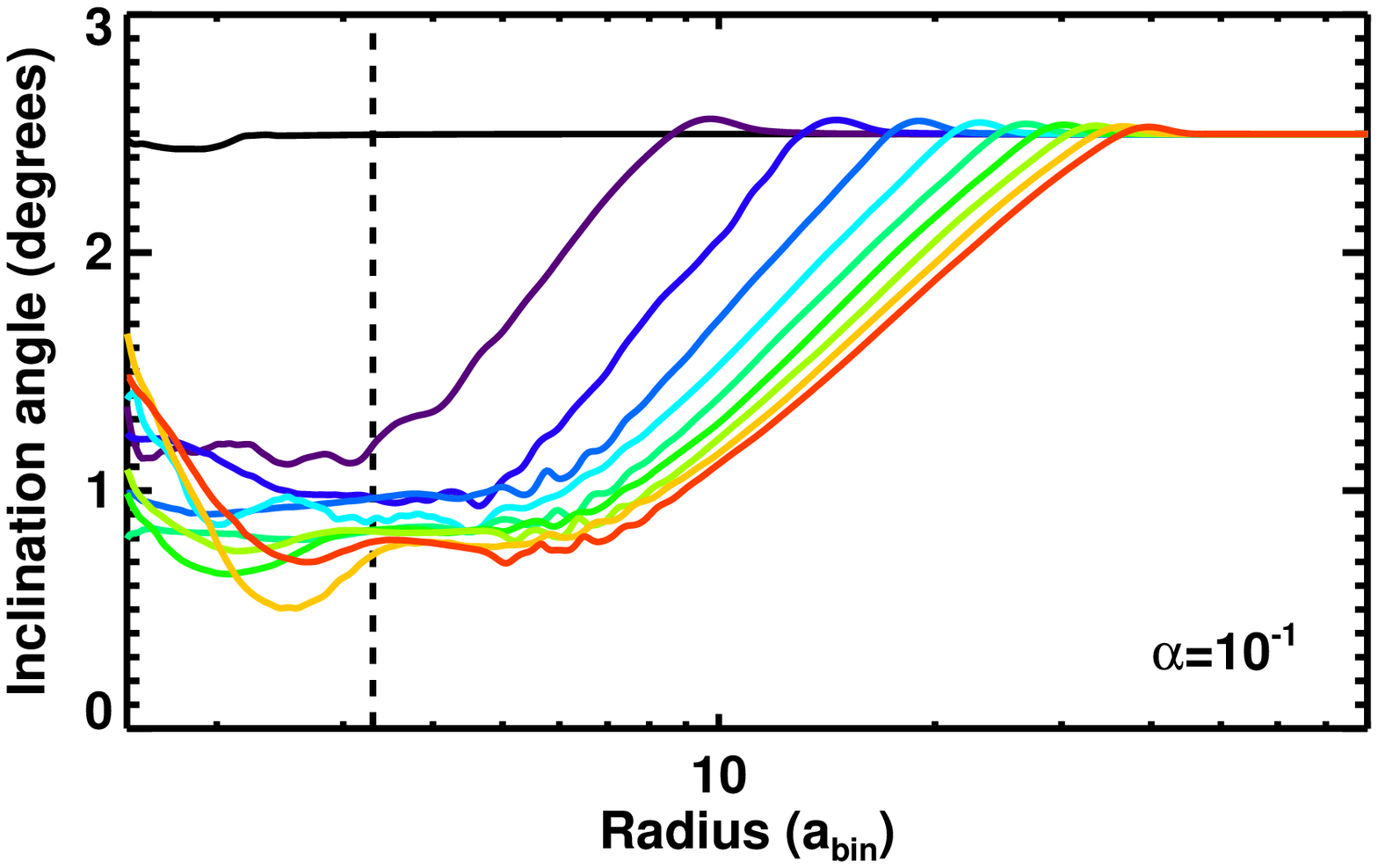}
\includegraphics[width=0.48\textwidth]{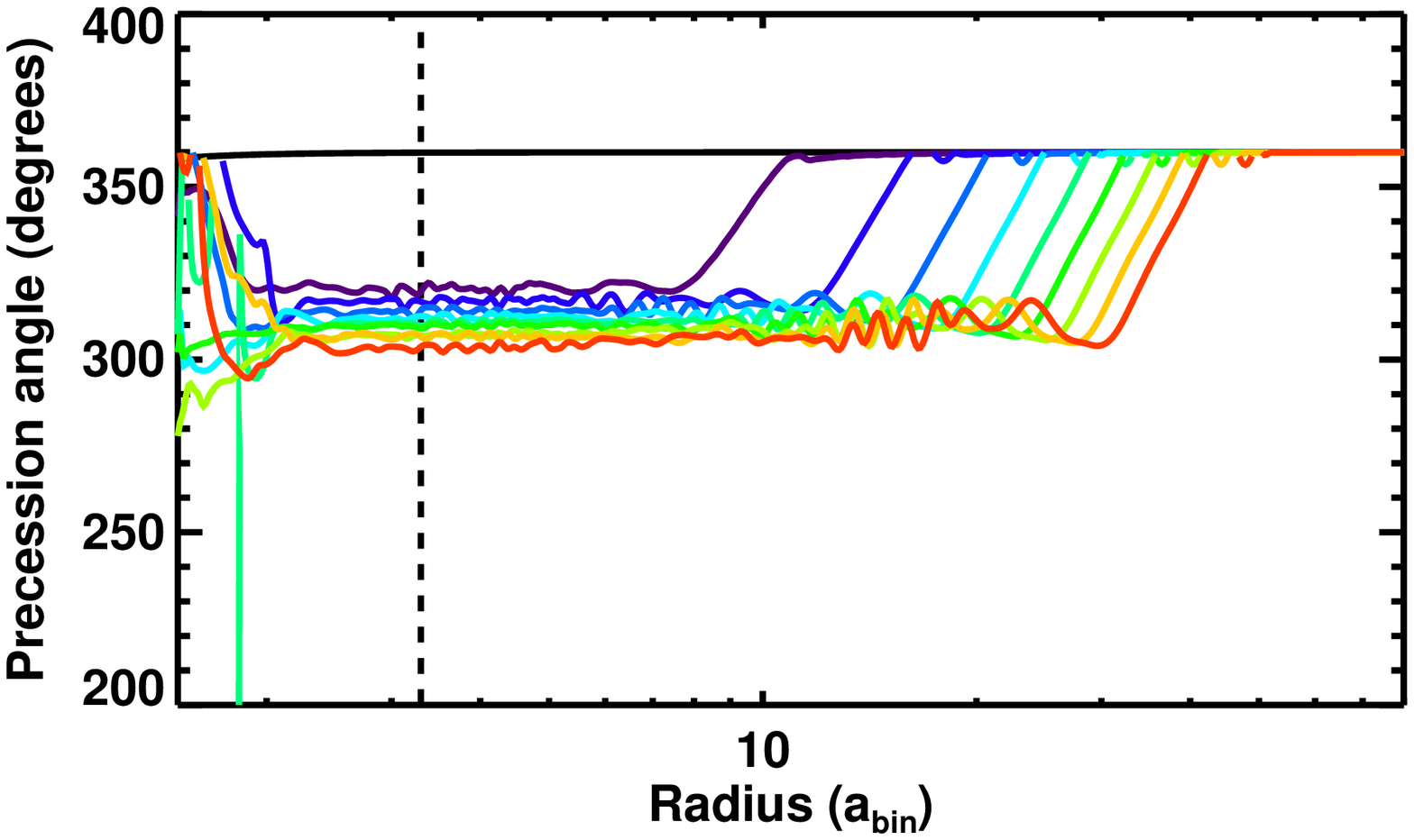}
\includegraphics[width=0.48\textwidth]{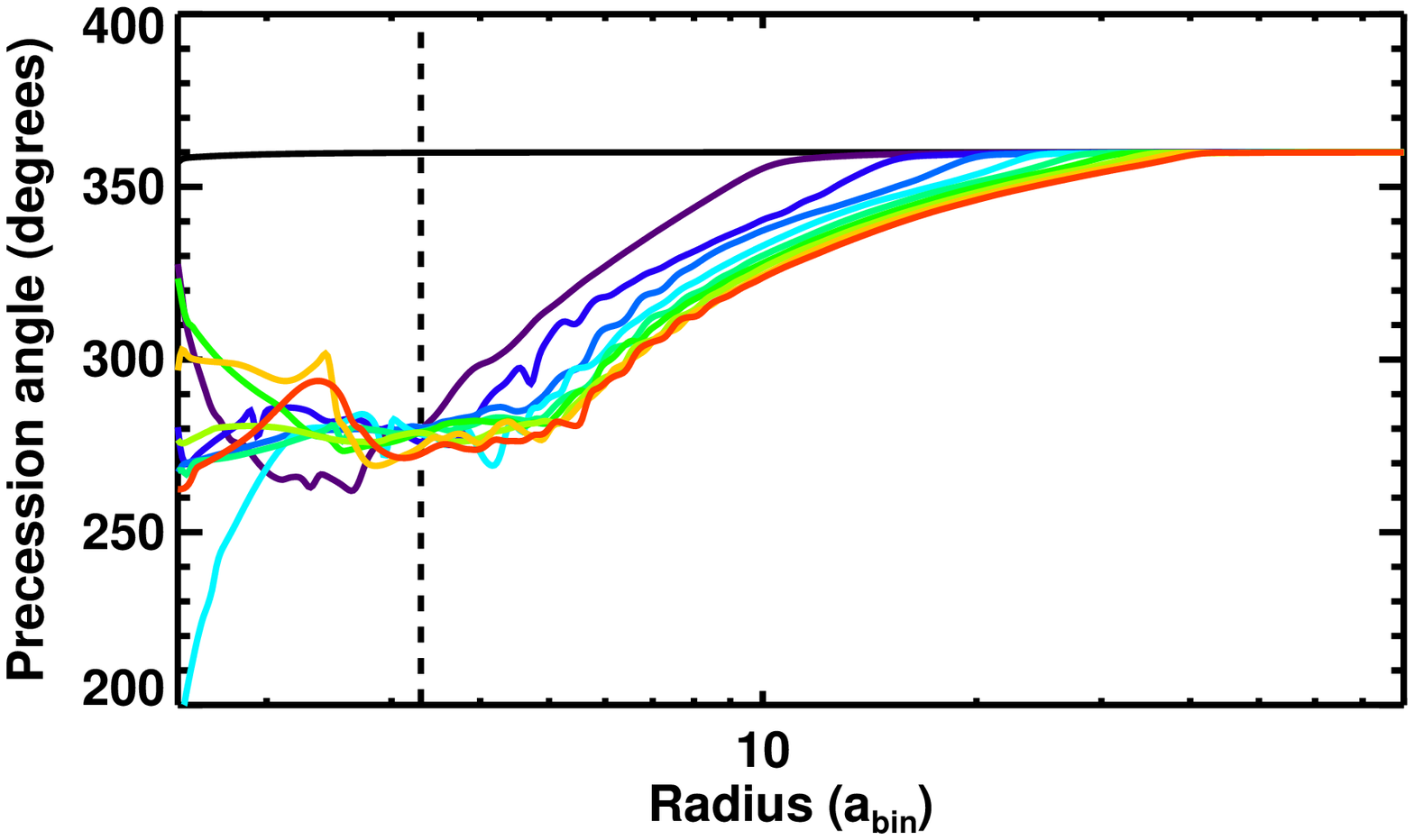}
\caption{{\it Left panel:} Disc inclination angle $i_D$ (top) and precession angle $\beta_D$ (bottom) as a function of radius at $t=1000$ $T_{bin}$, namely once a steady state is reached at the current location of Kepler-413b (vertical dashed line), for Model $5$ which has $\alpha=10^{-4}$. {\it Right:} same but for Model $6$ which has $\alpha=0.1$.}
\label{fig:models56}
\end{figure*}

Models $5$ and $6$ both have $h=0.05$ and $\gamma_{bin}=2.5^\circ$, but  the viscous stress parameter $\alpha$ is different in each model. Model $5$  has $\alpha=10^{-4}$  such that warp propagation occurs via bending waves, whereas Model $6$ has $\alpha=0.1$ so warps propagate diffusively because $\alpha>h$. The disc inclination and precession angles are shown in Fig. \ref{fig:models56}.  Model $5$ develops a very small warp  whose  structure at the end of the simulation is very similar to that of Model $2$ (see Fig. \ref{fig:model23}). This  is not surprising given that both models are in the wave regime with the same value for  the warp wave speed  $\sim c_s/2$.  As for Model $5$, the disc in Model 6 which has a larger viscosity also attains a state with almost no global precession at the end of the simulation,  but is found to be much more twisted, with a value for the twist between the inner and outer parts of the disc of $\sim 90$ degrees. This is 
in good agreement with the results of Fragner \& Nelson (2010) who studied discs in external binary systems, and found that discs with large viscosity tend to develop significant twists before achieving a state of solid 
body precession. As demonstrated by these authors, this arises because when warp communication is less efficient, the disc needs to be significantly distorted before internal stresses  are able to  counterbalance the effect of the differential precession induced by the binary.  Compared to Models 2 and 5, the amplitude of the warp established in Model 6 is also slightly higher, with the inner and outer regions of the disc having relative inclination of $\sim 1.8$ degrees.  This can be also observed in Fig. \ref{fig:warp3d} which shows column density plots for a particular viewing geometry for Models $5$ and $6$ at $t=700$ $T_{bin}$. For both models,  we would again expect alignment between the disc and binary orbital plane to occur on a time corresponding to  the global viscous evolution timescale of the disc once a steady warped structure has been established, corresponding to $\sim 4.5\times 10^8 $ and $\sim 4.5\times 10^5 $ $T_{bin}$ for models $5$ and $6$, respectively.

\label{sec:results}

\subsection{Evolution for a flared disc model: Model 7}
\label{sec:model7}
\begin{figure}
\centering
\includegraphics[width=0.48\textwidth]{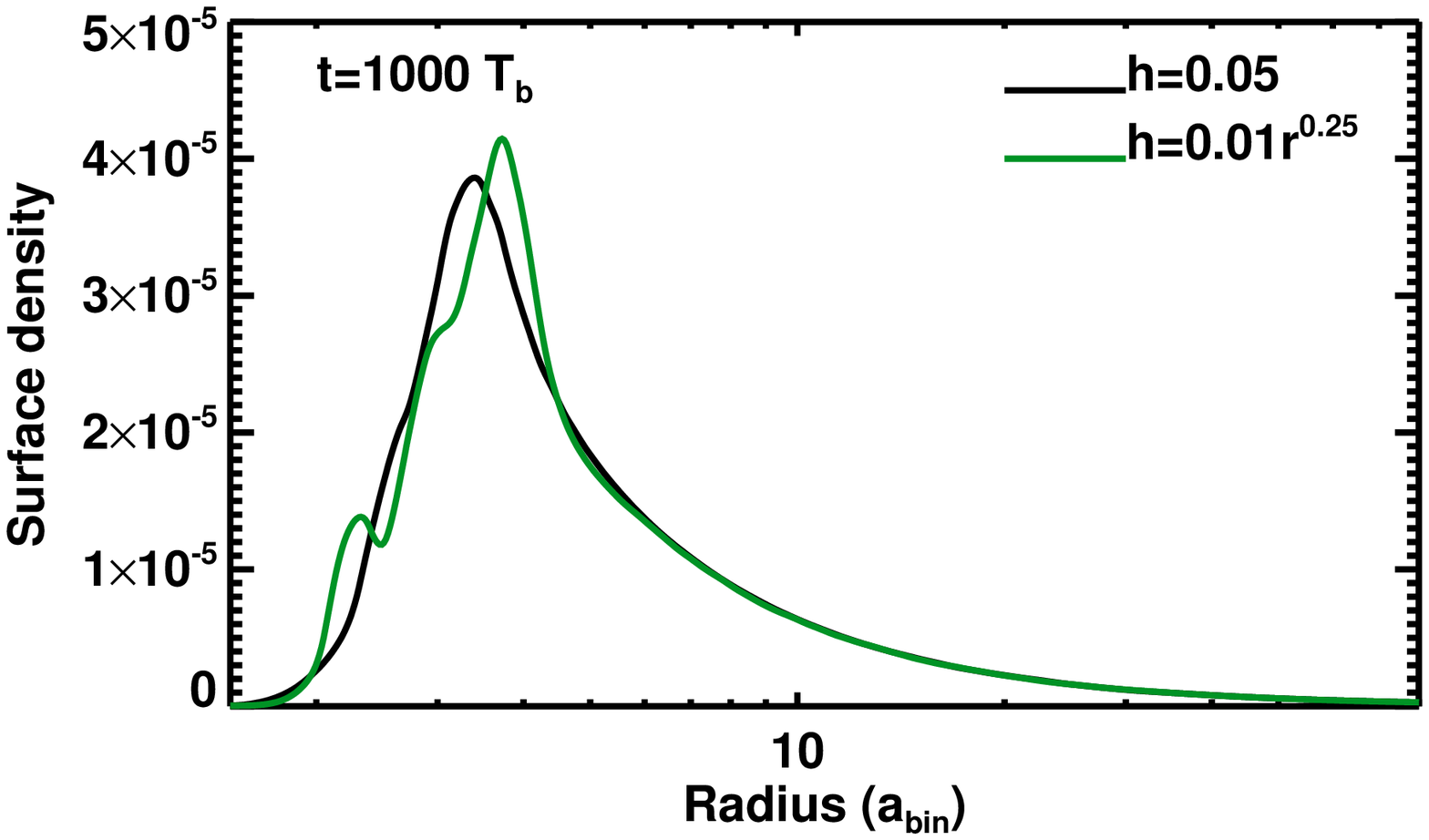}
\includegraphics[width=0.48\textwidth]{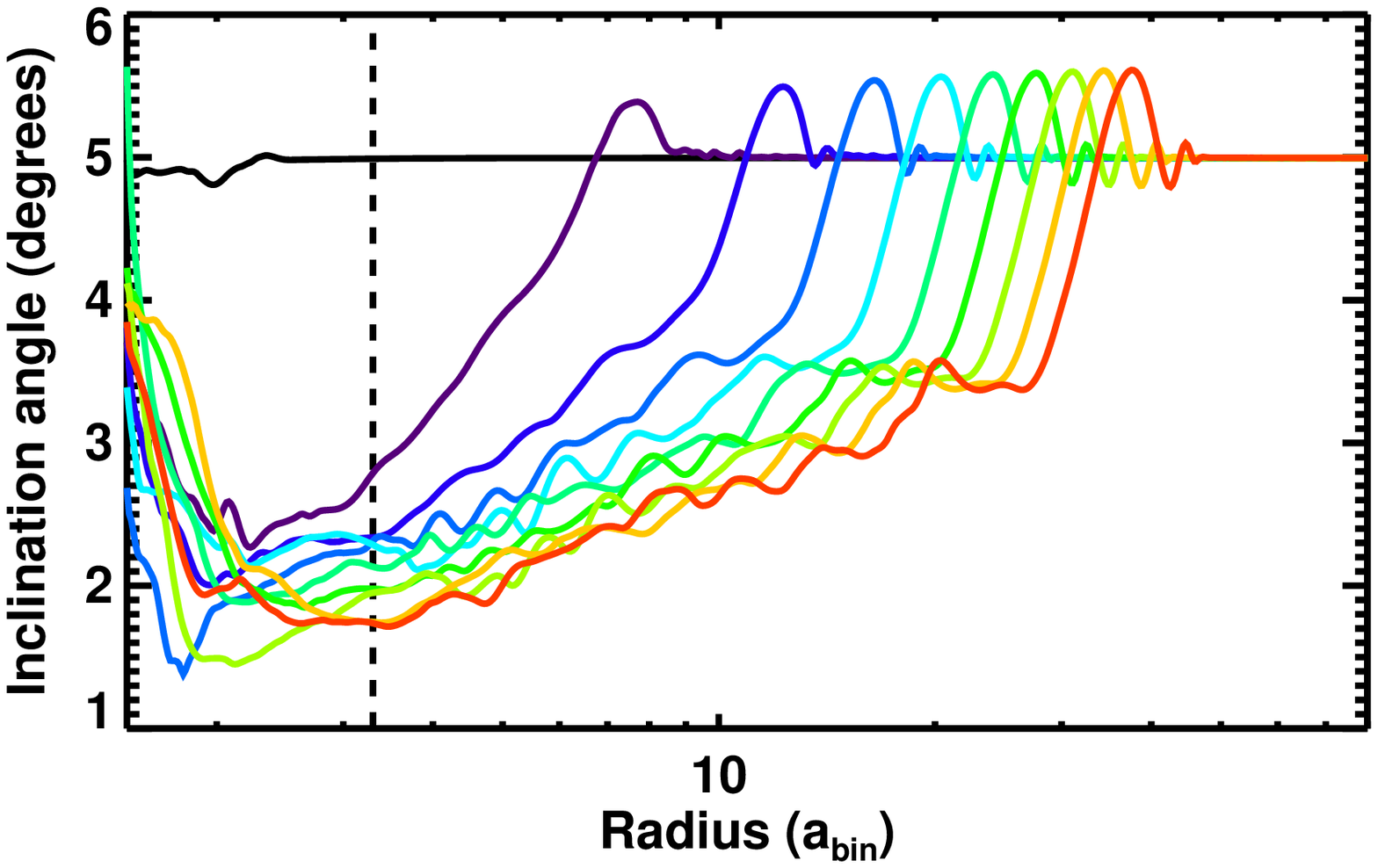}
\includegraphics[width=0.48\textwidth]{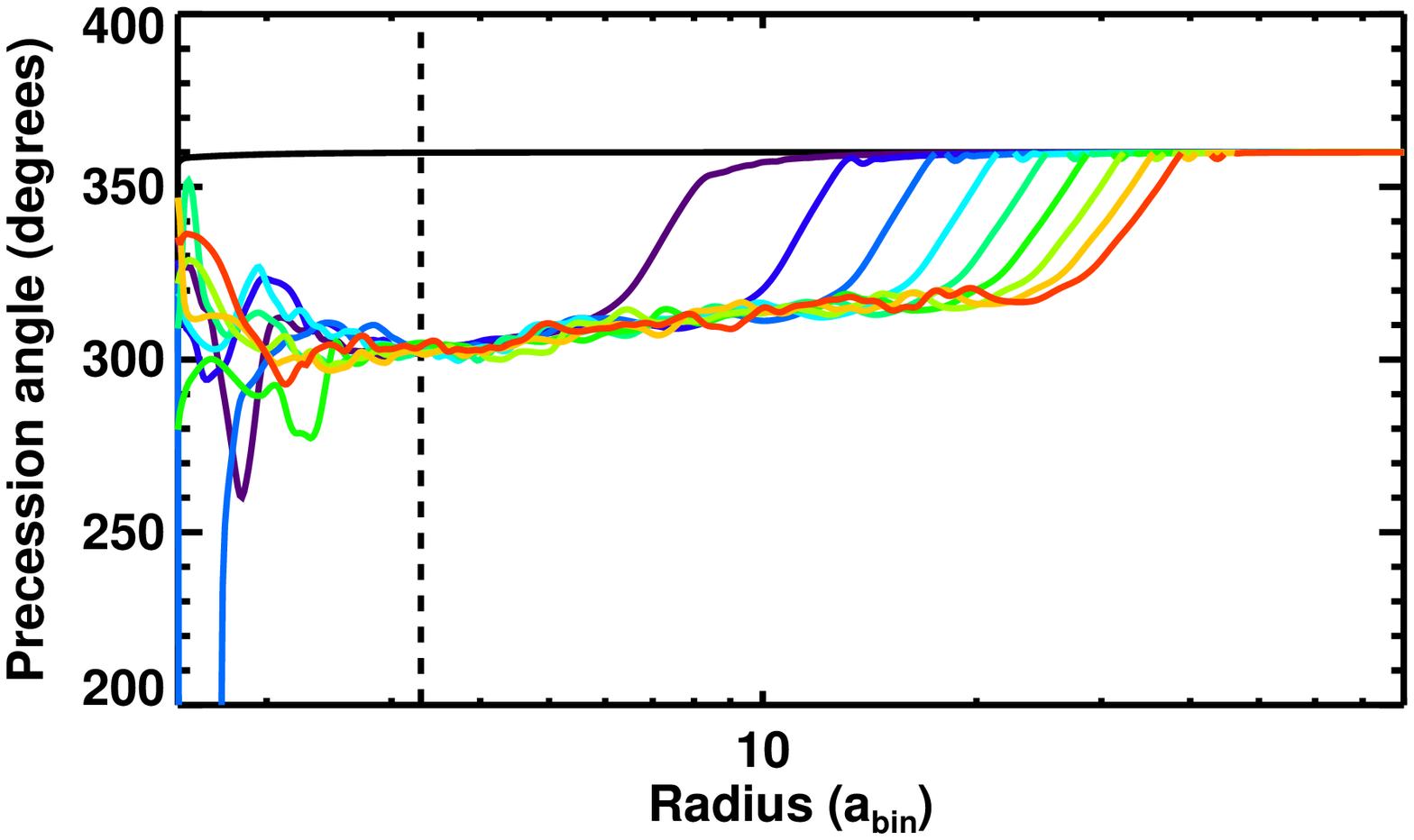}
\caption{{\it Upper panel:} Surface density profile at $t=1000$ $T_{bin}$, namely once a steady state is reached at the current location of Kepler-413b (vertical dashed line), for Models $1$ and $7$. Model $1$ has constant aspect ratio 
with $h=0.05$  whereas Model $7$ corresponds to a flared disc with $h=0.0(r/r_0)r^{0.25}$. {\it Middle panel:} Disc inclination angle $i_D$ as a function of radius for Model $7$ at times $t=1$ (left) $, 100, 200,..., 1000$ (right) $T_{bin}$.  {\it Lower panel:} Disc precession angle $\beta_D$ as a function of radius for Model $7$ at the same times. }
\label{fig:model7}
\end{figure}

We now discuss the circumbinary disc evolution for Model $7$, which corresponds to a flared disc with aspect ratio $h=0.01(r/r_0)^{0.25}$. The aspect ratio is clearly smaller than for Model 1, such that we expect the tidally truncated inner cavity to be deeper for Model 7 (e.g. Pierens \& Nelson 2013). The surface density profile at $t=1000$ $T_{bin}$ is shown in the upper panel of Fig. \ref{fig:model7}.  The density peak is located slightly further out at $R\sim 3.8$ for this run compared to Model 1, consistent with expectations, and the cavity edge is clearly steeper.

 Using Eq . \ref{eq:omegap} for the free particle precession rate, we  estimate $t_w<t_{prec}=\Omega_{prec}^{-1}$ in regions where $r\gtrsim 6$ $a_{bin}$, which is slightly larger than the tidal truncation radius. The implication is that we expect a small amount of warping to be established in the innermost disc regions.  For this model, the disc inclination and precession angles are plotted as functions of time and radius in the middle and lower panels of Fig. \ref{fig:model7}, respectively. Similarly to the previous models that we have considered, a quasi-steady configuration is reached at $t\sim 1000$ $T_{bin}$ interior to $r\sim 40\; a_{bin}$.  This quasi-steady state is also characterised by a small degree of warping near $R\sim 5-6$ $a_{bin}$ and a small twist, as expected. At $t=1000$ $T_{bin}$, the values at $R=3.5\; a_{bin}$ for the disc inclination and precession angles are  $i_D\sim 1.7^\circ$ and $\beta\sim 300^\circ$ respectively, which  are close to those that we have obtained in Model 1. The outer edge of where the disc is being distorted is also very similar to what has been found in Model 1. This confirms that this location is set by differential precession rather than by bending waves, since these propagate much  slower in Model 7.  \\
The main 
 difference here will be that a planet with the same mass as Kepler-413b will open a deep gap in the Model 7 disc, whereas it will at most create a small density depression and  experience Type I migration in Model 1. In the next section, we discuss the dynamical evolution of planets that are initially \emph{locally} aligned with the circumbinary disc.

\section{Dynamics of planets embedded in warped circumbinary discs}
\subsection{Evolution for two fiducial cases: Models $1$ and $7$}
To examine the evolution of planets that are initially locally aligned with the circumbinary disc, we focussed on the Models $1$ and $7$, which have $h=0.05$ and $h=0.01(r/r_0)^{0.25}$, respectively. The viscous stress parameter $\alpha=4\times 10^{-3}$ in both models. We inserted a non-accreting planet with mass ratio 
 $q \equiv m_p/M_{bin}=1.6\times 10^{-4}$ and semi-major axis $a_p=3.5$ $a_{bin}$ on a circular orbit after each of the discs had evolved for $1000$ $T_{bin}$. Hence, the planet is located close to the tidal truncation radius at the beginning of this phase of the simulations.
 
\begin{figure*}
\centering
\includegraphics[width=0.48\textwidth]{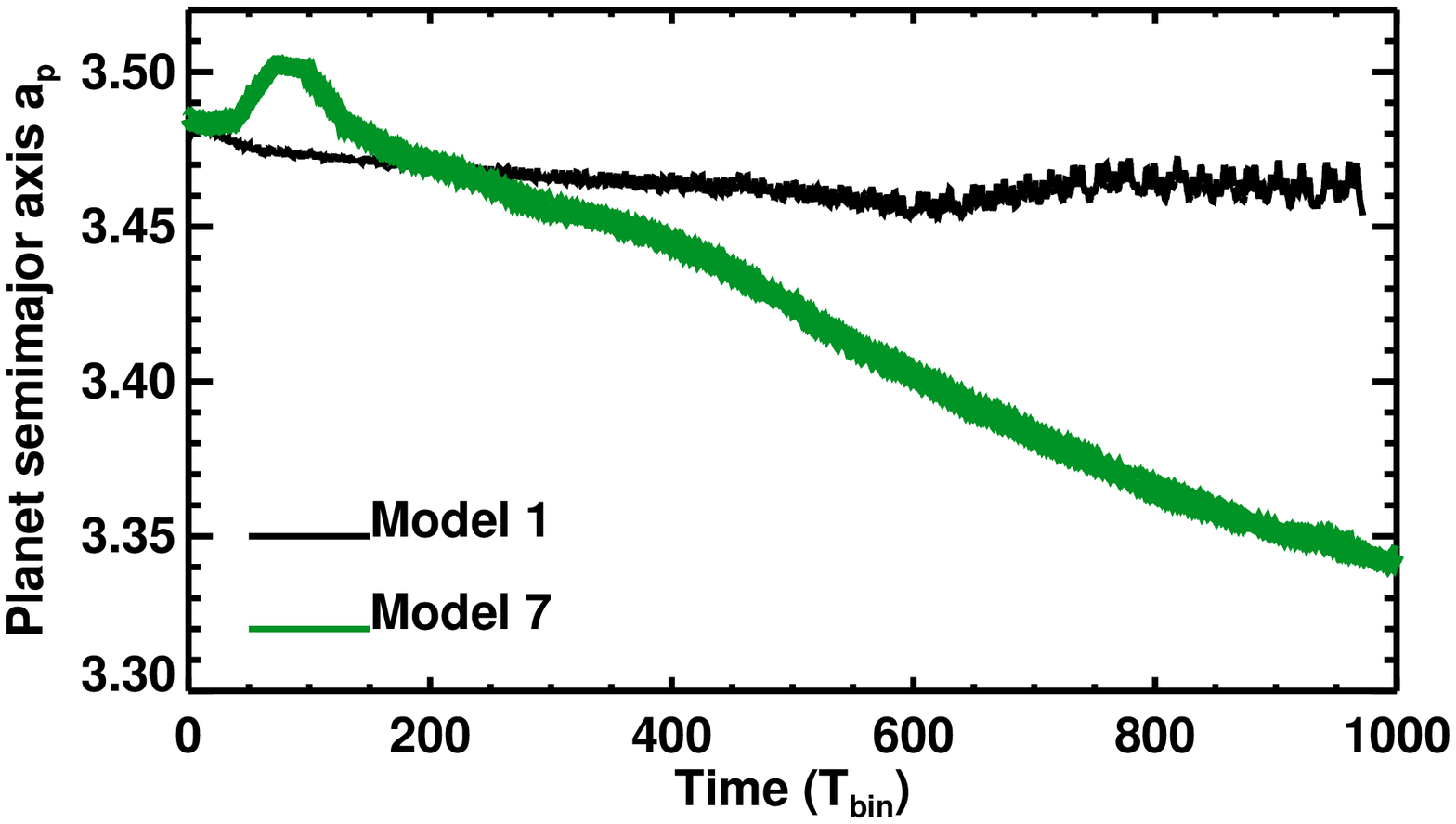}
\includegraphics[width=0.48\textwidth]{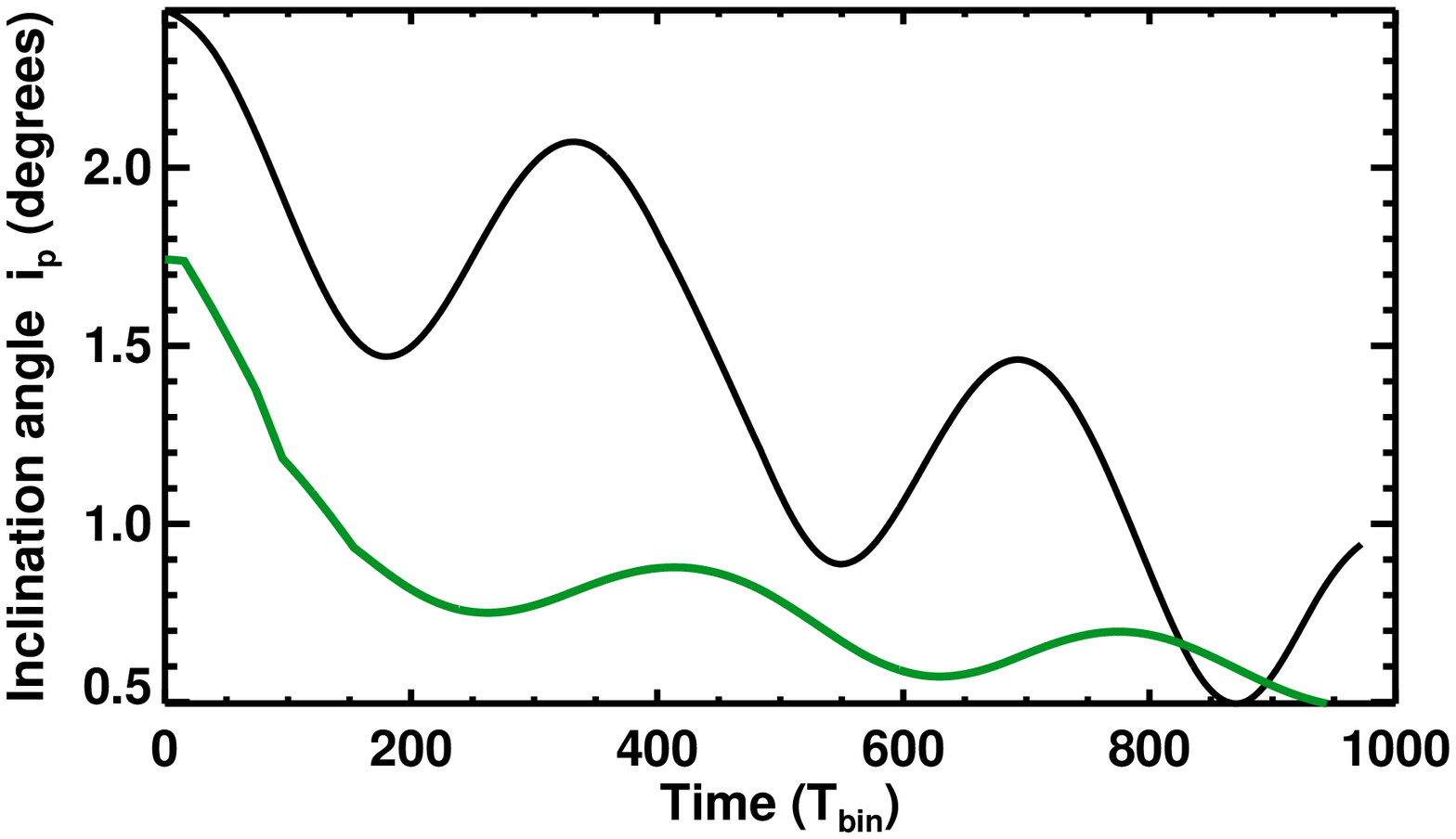}
\includegraphics[width=0.48\textwidth]{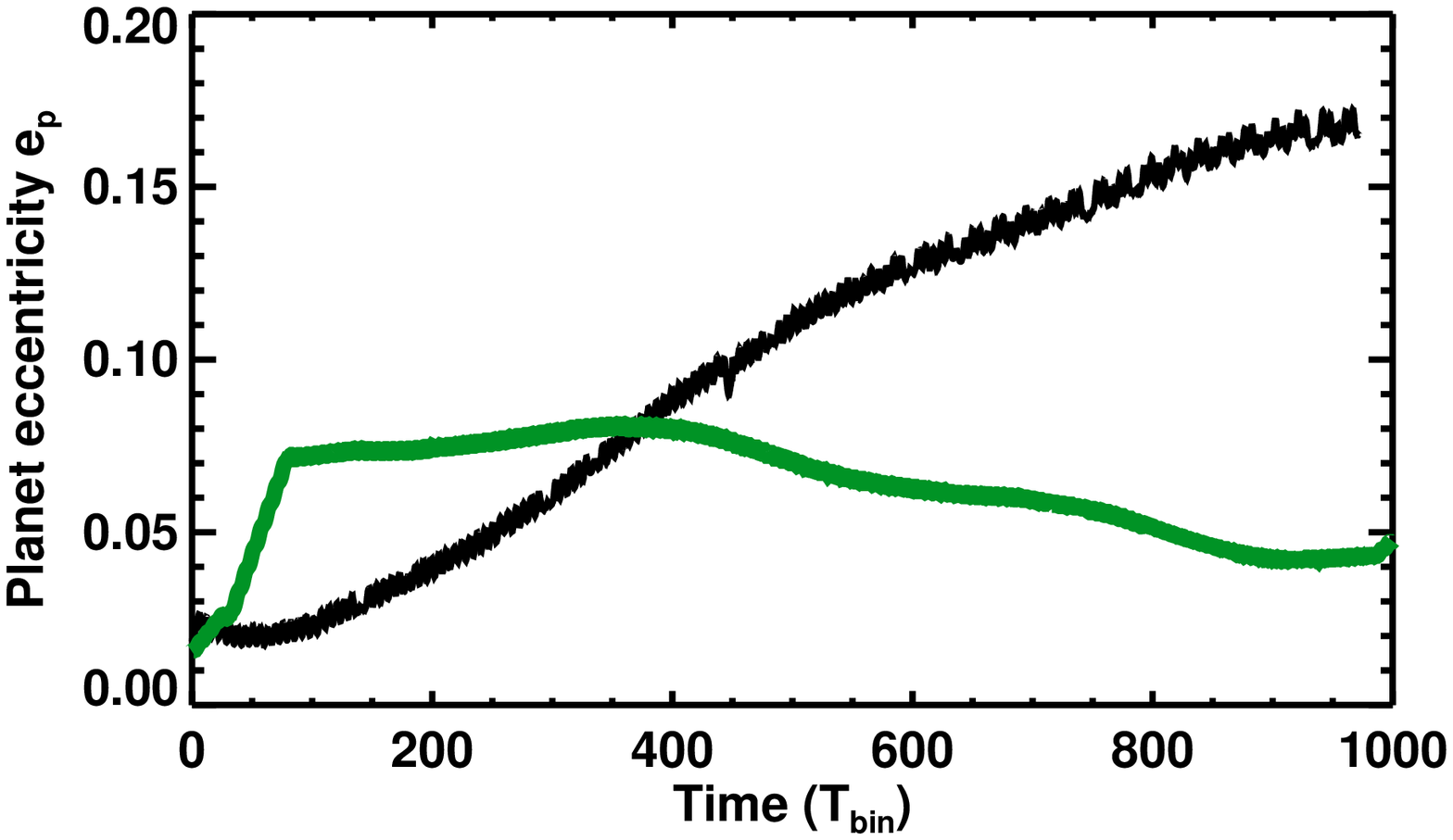}
\includegraphics[width=0.48\textwidth]{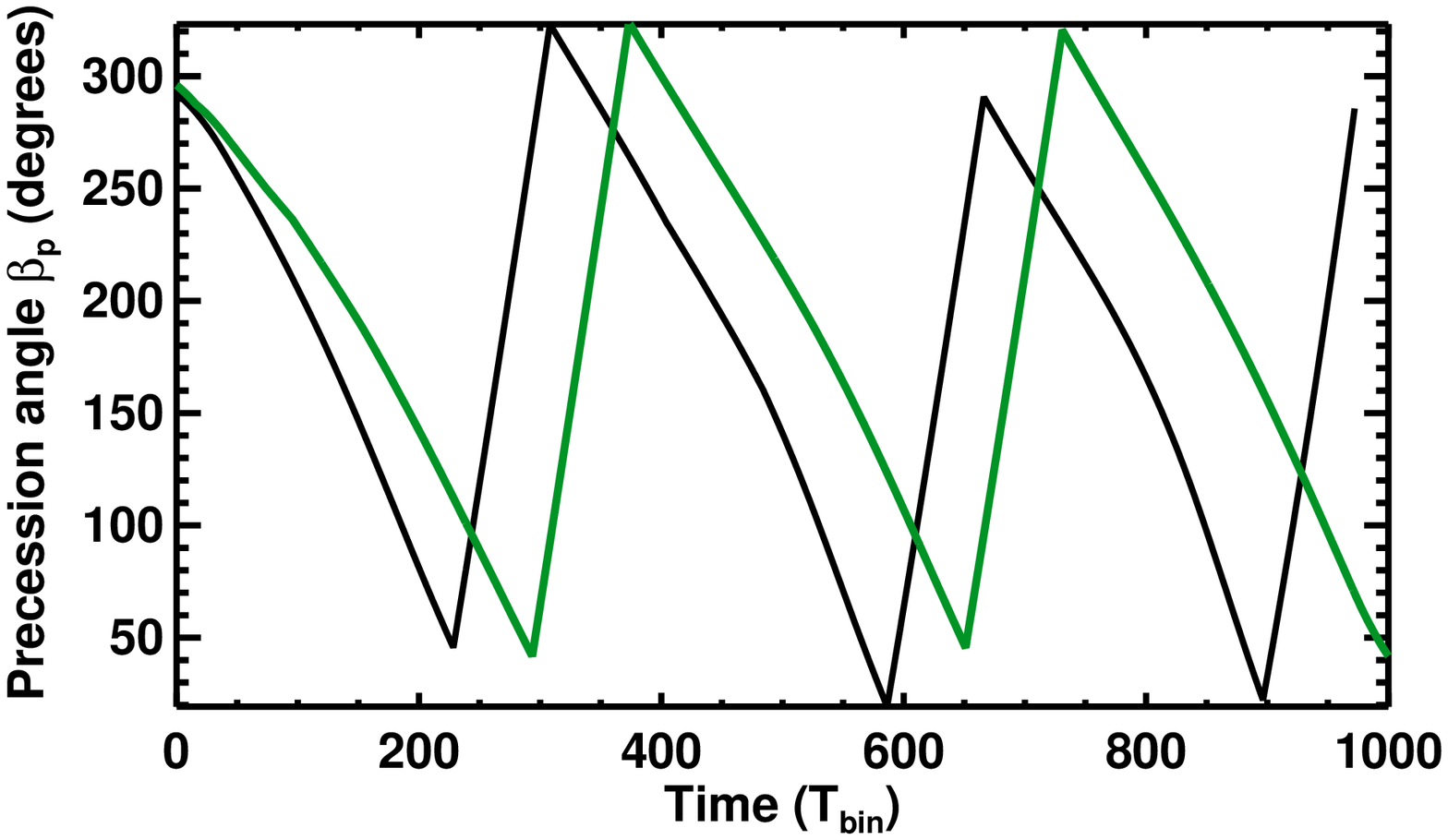}
\caption{{\it Left panel:} Time evolution of the planet semi-major axis (top) and eccentricity (bottom) for Models $1$ and $7$. 
The planet undergoes Type I migration for Model $1$ whereas it opens a gap and undergoes Type II migration for Model $7$. {\it Right panel:} Time evolution of the planet inclination $i_p$ (top) and precession angle $\beta_p$ (bottom) for the same models.  }
\label{fig:1mmsn}
\end{figure*}

\begin{figure}
\centering
\includegraphics[width=0.48\textwidth]{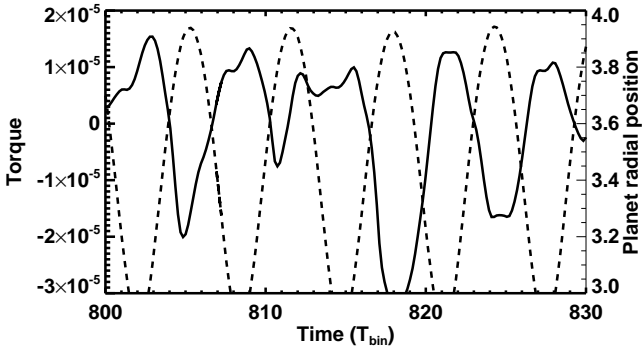}
\caption{Evolution of the orbital radius (dashed line) of Kepler-413b and evolution of the specific torque (solid line) experienced by the planet over a few orbital periods for Model $1$.}
\label{fig:torque}
\end{figure}

We begin discussion of the simulation results by considering discs with masses equivalent to $1$ MMSN. We discuss how the results depend  on disc mass in the next section. The left panel of Fig. \ref{fig:1mmsn} shows the evolution of the planet semi-major axis and eccentricity for the two models. The criterion for deep gap formation (Crida et al. 2006) is
\begin{equation}
{\cal P}=1.1\left(\frac{q}{h^3}\right)^{-1/3}+\frac{50\alpha h^2}{q}\le 1
\label{eqn:Crida}
\end{equation}
For Model 1, ${\cal P} \sim 4$ such that gap formation should not occur and migration will be in the Type I regime. In that case, previous work has shown that the planet can become trapped at the edge of the inner cavity for two different reasons: i) halting of migration through the action of corotation torques in a region where the surface density increases with radius (Pierens \& Nelson 2007); ii) stopping of inward migration once the planet eccentricity becomes high enough to induce a Lindblad torque reversal (Pierens \& Nelson 2013). Inspection of Fig. \ref{fig:1mmsn} reveals that migration is halted at the cavity edge, as expected. The horseshoe libration timescale $t_{lib}=8\pi a_p /3 \Omega_p x_s$ is significantly shorter than the viscous timescale across the horseshoe region $t_\nu=x_s^2/\nu$ , where $x_s=1.2\sqrt{q/h^3}$ is the  half-width of the horseshoe region (Paardekooper et al. 2010), such that corotation torques are partially saturated and therefore significantly attenuated. The implication is that stalling of migration likely occurs because of Lindblad torque reversal. To explain this effect, Pierens \& Nelson (2013) showed that a planet on a significantly eccentric orbit has smaller angular velocity at apocentre than neighbouring disc material, resulting in the outer disc exerting a positive torque on the planet. The opposite happens at pericentre where the inner disc exerts a negative torque. Hence, the torque on the planet oscillates from negative to positive values as the planet moves from apocentre to pericentre and back again. The  time averaged total torque, however, can equal zero resulting in migration being halted.  Figure~\ref{fig:torque} displays the total torque together with the planet radial position over a few orbital periods, and shows that the torque is indeed positive (resp. negative) when the planet is at apocentre (resp. pericentre), but with the orbit-averaged torque almost cancelling as expected for Lindblad torque reversal. As noted by Pierens \& Nelson (2008), the slight phase shift between the two curves is due to the time needed for the planet to create an inner (resp. outer) wake in the disc at apocentre (resp. pericentre). 
 
\begin{figure}
\centering
\includegraphics[width=0.48\textwidth]{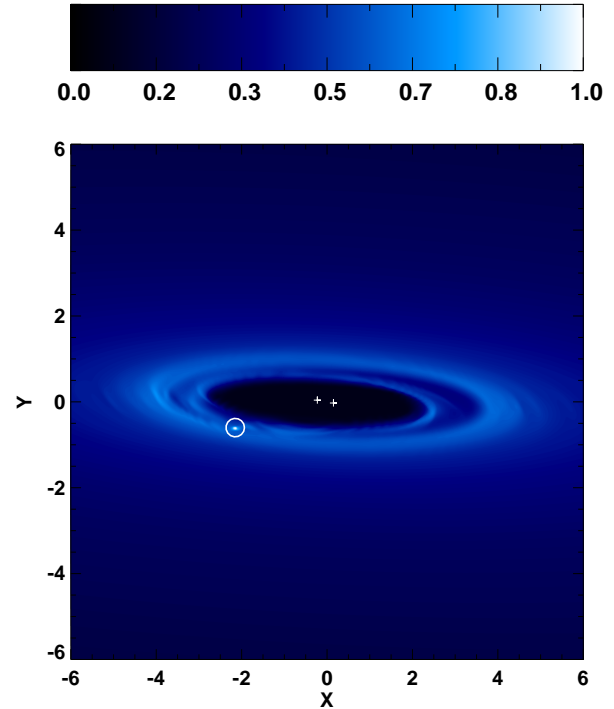}
\caption{{\it Upper panel:} Projected disc surface density along the line of sight characterized by the angles $(\theta,\phi)=(70^\circ, 80^\circ)$ for Model $7$ at $t=1000$ $T_{bin}$. }
\label{fig:flared2d}
\end{figure}

 Model $7$ resulted in the planet migrating inwards over the duration of the run.  This is because the planet opens a gap in the disc. The gap opening criterion given by Eq.~(\ref{eqn:Crida}) gives ${\cal P} \sim 0.6$, such that deep gap formation is expected, and is confirmed by   inspection of Fig. \ref{fig:flared2d} which shows a projection of the column density through the disc for a viewing angle $(\theta,\phi)=(70^\circ,80^\circ)$ at time $t=1000$ $T_{bin}$ after the planet was inserted. The inner disc is depleted so inward migration is driven by the outer disc, pushing the planet into the cavity created by the binary.  Due to the very long run times required by the simulations, the final fate of the system remains uncertain. It is possible that the planet will approach $6:1$ mean motion resonance with the binary at later times. Previous work has shown different outcomes for long term simulations involving gap forming planets. These include ejection of the planet from the system through interaction with the binary, scattering into the outer disc by the binary such that the planet repeatedly migrates to the inner cavity and is scattered out again, and slow inwards migration that stalls when the density of gas in the cavity becomes very small (Nelson 2003; Pierens \& Nelson 2008; Pierens \& Nelson 2013).
 
We notice that for the two models that we have considered here, the planet eccentricity reaches a constant value but we do not obtain particularly good agreement with observational data. Model $1$ overestimates the value for $e_p$ ($e_p\sim 0.15$ at the end of the run) whereas the value 
for $e_p$ inferred from Model $7$ ($e_p\sim 0.05$) is much smaller that the observed value $e_p\sim 0.1$. We note that high values of the eccentricities are generally driven by non-axisymmetric features in the disc and not by the central binary in these types of simulations (Kley \& Nelson 2010). Presumably a model with parameters somewhere between those of Model 1 and Model 7 could produce a better fit.
 
The right panel of Fig. \ref{fig:1mmsn} shows the time evolution of the planet inclination $i_p$ and precession angle  $\beta_p$ of the planet angular momentum vector around the angular momentum of the binary. These are defined by: 
\begin{equation}
\cos(i_p)=\frac{{\bf L_p}\cdot {\bf L_B}}{|{\bf L_B}||{\bf L_p}|}
\end{equation}
and
 \begin{equation}
 \cos(\beta_p)=\frac{{\bf L_p}\times {\bf L_B}}{|{\bf L_B}\times {\bf L_p}|}\cdot {\bf u}
 \end{equation}
where ${\bf L_p}$ is the orbital angular momentum vector of the planet. 
Since the disc angular 
momentum is much higher than that of the planet, changes in the global disc inclination resulting from 
secular interaction with the planet is expected to be small. Although not shown here, we checked that the 
disc inclination $i_D$ and precession angles $\beta_D$ remain almost constant over the duration of the simulation ($\sim 1000$ $T_{bin}$). Looking at the right panel of Fig. \ref{fig:1mmsn}, it is immediately obvious that coplanarity between the disc and planet is not maintained for a $1$ MMSN disc,
and the planet orbit tends to align with the binary orbit. Since $(i_p-i_d)^2 \propto (\beta_p-\beta_d)^2$ (Fragner, Nelson \& Kley 2011), the nodal precession of the planet induced by the binary, which occurs significantly faster than the disc precession, induces the relative inclination between the planet and local disc midplane to grow at early times,  leading to the initial 
decrease in $i_p$ that can be observed in the upper right panel of Fig. \ref{fig:1mmsn}. $i_p$ then oscillates with period $T_{osc}\sim 300$ $T_{bin}$ which, for $r\sim3.4$, is close to the free particle precession period $2\pi/\Omega_{prec}(r)$, where $\Omega_{prec}(r)$ is given by Eq. \ref{eq:omegap}. As expected, the relative tilt $i_p-i_d$ (not shown here) is at a minimum when the disc and planet precession angles are in phase whereas it is at a maximum when $|\beta_p-\beta_d |=\pi$. For both Models 1 and 7, the long term evolution involves the planet inclination relative to the binary progressively decreasing. This occurs because the binary causes the planet to continuously precess relative to the disc. When the planet and disc planes are misaligned the passage of the planet through the disc occurs at high velocity, and the disc provides a source of dynamical friction on the planet's orbit. In the limit that the rapid precession of the planet is driven by the binary, with the disc only making a negligible contribution by virtue of it having a small mass, the binary orbit plane provides a plane of symmetry to which the planetary orbit damps due to the disc-induced dynamical friction.



\label{sec:results}
 
\label{sec:results}

\subsection{Dependence on disc mass}
To examine how the disc mass $M_D$ affects the inclination evolution of embedded planets, we performed two additional Model 1 simulations with disc masses equivalent to $M_D=2$ and 20 MMSN.  We note in passing that the model with disc mass equivalent to $20$ MMSN is used for illustrative purpose only, since it would be gravitationally unstable outside 
$R\sim 60$ $a_{bin}$. Fig. \ref{fig:1mmsn} shows the planet inclination and precession angles as a function of time for $M_D=1$, 2, and 20 MMSN.  As expected, the planet tries to maintain coplanarity with the disc as the disc mass is increased, since the gravitational acceleration of the disc tries to pull the planet back towards the midplane, in competition with the binary which tries to drive precession of the planet relative to the disc, and hence away from the disc midplane. For $M_D\ge 2$ MMSN, we see that the inclination and precession angle exhibit damped oscillations, leading $i_p$ to reach a constant value that increases with disc mass. For $M_D=2$ MMSN, $i_p\sim 1.2^\circ$ once a quasi steady state is reached whereas for $M_D=20$ MMSN, the planet remains almost fully aligned with the disc.  For both cases, the precession angle is also observed to librate around a value that is very close to that of the disc (see Fig.~\ref{fig:model1}),  whereas it circulates in the case with $M_D=1$ MMSN. In summary, our results show that in the limit of high disc mass, the precession of the planet relative to the disc is prevented from occurring because disc gravity causes the planet to remain orbiting very close to the disc midplane. In a low mass disc, precession is completely dominated by the central binary, and the damping of the disc-planet relative motion as the planet passes through the disc during periods of high mutual inclination leads to the planet orbit damping towards the binary orbit plane. For intermediate disc masses, the planet orbit achieves a stationary state with a precession angle that is just offset from that of the non-precessing disc, and a non-zero orbital inclination that is smaller than that of the disc.

\subsection{A secular model with inclination damping, migration and disc dispersal}
In the context of secular theory, libration of the precession angle is expected for sufficiently massive discs that induce onto the planet a precessional torque that is stronger than the one induced by the binary. The inclination damping effect, however, is due to resonant planet-disc interactions which tend to damp the planet orbit into the local disc plane. For very high mutual inclinations, a model of inclination damping based on dynamical friction can also be formulated. 

The evolution of the disc-planet-binary system under the influence of secular gravitational interactions, migration and inclination damping can examined in a simplified formalism by considering the evolution equation for the planet tilt $W_p=l_{p,x}+i l_{p,y}$ including effects resulting from  i) secular interaction with both the binary and the disc, and ii) resonant interaction with the disc. Following Lubow \& Martin (2016), the secular evolution equation for the planet  is given by:
\begin{equation}
J_p\frac{dW_p}{dt}=i C_{pd}(W_d-W_p)-iC_{ps} W_p,
\end{equation}
where $J_p=m_pa_p^2\Omega_p$ is the planet angular momentum, $W_D=l_{d,x}+i l_{d,y}$ is the disc tilt,  and $C_{pd}$ (resp. $C_{ps}$) is a coupling coefficient between the planet and the disc (resp. binary).  These are given by:
\begin{equation}
C_{pd}=2\pi \int_{R_{in}}^{R_{out}}Gm_pR\Sigma(R)K(R,a_p)dR
\label{eq:cpd}
\end{equation}
and
\begin{equation}
C_{ps}=Gm_pM_2K(a_p,a_{bin}),
\end{equation}
where the kernel $K$ is given by:
\begin{equation}
K(R_i,R_j)=\frac{R_iR_j}{4\pi}\int_{0}^{2\pi}\frac{\cos(\phi)d\phi}{(R_i^2+R_j^2-2R_iR_j\cos(\phi))^{3/2}}.
\end{equation}
Regarding the evolution equation for the disc, we only include the term corresponding to the secular interaction with the planet so that it reads: 
\begin{equation}
J_p\frac{dW_d}{dt}=i C_{pd}(W_p-W_d)
\end{equation}
where
\begin{equation}
J_d=2\pi\int_{R_{in}}^{R_{out}} \Sigma(R)R^3\Omega(R) dR
\end{equation}
is the disc angular momentum. The effect of the planet on the disc is expected to be small because of the small angular momentum of the planet compared to the disc, but is included for completeness. The effect of the binary on a disc with large radius is also negligible in terms of driving its precession, as demonstrated by our simulations.

The effect of the resonant interaction with the disc 
can be incorporated by adding an additional term  (Lubow \& Ogilvie 2001) to the secular evolution equation for the planet, which then becomes:
\begin{equation}
J_p\frac{dW_p}{dt}=i C_{pd}(W_d-W_p)-iC_{ps} W_p-J_p\frac{W_p-W_d}{\tau_{inc}}
\end{equation}
where $\tau_{inc}$ is the inclination damping timescale which is set to (Tanaka \& Ward 2004):
\begin{equation}
\tau_{inc}=\frac{t_{wave}}{0.544}
\end{equation}
with:
\begin{equation}
t_{wave}=q^{-1}\frac{\Sigma_p a_p^2}{M}h^4\Omega_p^{-1}
\end{equation}
where $\Sigma_p$ is the disc surface density at the location of the planet. We solved the secular evolution equations for the disc and planet,  assuming that the disc and planet are initially coplanar 
with $i_p=i_d=2.5^\circ$ and $\beta_p=\beta_d=0^\circ$. We  adopted disc parameters corresponding to the initial disc parameters for Model $1$, and used the surface density profile at equilibrium to compute the initial planet--disc coupling coefficient $C_{pd}$ in Eq. \ref{eq:cpd}. To model gas disc dissipation due to accretion onto the star and photoevaporation, this coefficient and the disc angular momentum $J_D$ are forced to decay exponentially with an e-folding time $t_{dis}$, for which we adopt a reference value of  $10^5 T_{bin}$. The planet is also allowed to migrate on a timescale $t_m=J_p/2\dot{J}_p$, where ${\dot J}_p$ is the disc torque acting on the planet which is given by the sum of the Lindblad torque and fully unsaturated horseshoe drag. In that case, $\dot{J}_p$ is given by (Paardekooper et al. 2010):
\begin{multline}
{\dot J}_p/\Gamma_0=-(2.5-0.5\beta-0.1\sigma)\left(\frac{0.4}{b/h}\right)^{0.71}-1.4\beta\left(\frac{0.4}{b/h}\right)^{1.26}+\\
1.1
\left(\frac{3}{2}-\sigma\right)\left(\frac{0.4}{b/h}\right)
\end{multline}
where $\beta=1$ and $\sigma$ are the negatives of the power-law indices for the temperature and surface density, respectively, and where:
\begin{equation}
\Gamma_0=(q/h)^2\Sigma_pa_p^4\Omega_p^2
\end{equation}

\begin{figure}
\centering
\includegraphics[width=\columnwidth]{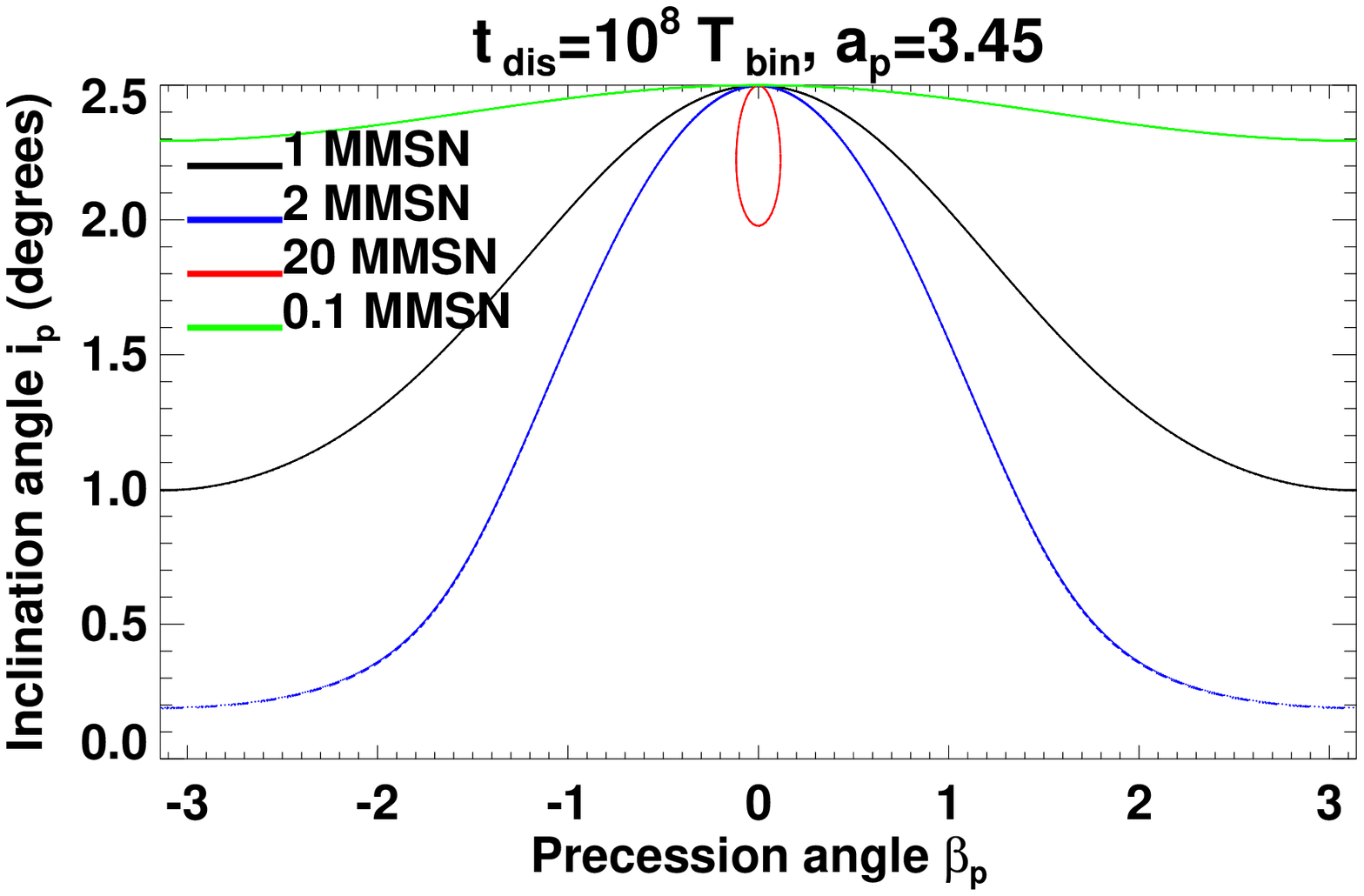}
\includegraphics[width=\columnwidth]{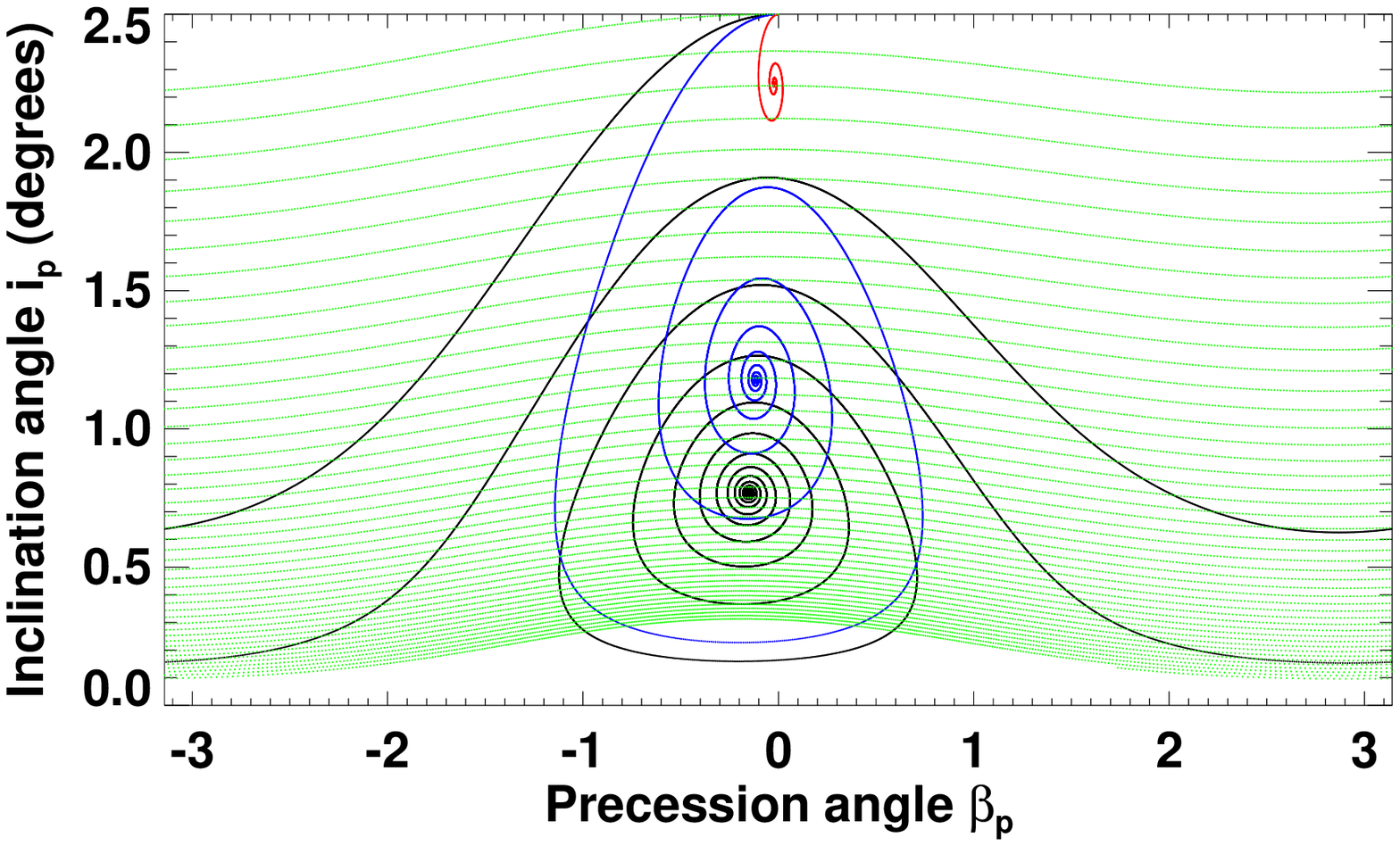}
\caption{{\it Upper panel:} Planet inclination $i_p$ as a function of the planet precession angle $\beta_p$, namely phase portrait  obtained from a secular evolution model in which the effect of the resonant interaction with the protoplanetary disc is not included. Employed disc parameters correspond to those of Model $1$. {\it Lower panel:} same but in the case where the effect of the resonant interaction with the disc is included. }
\label{fig:secular}
\end{figure}

In order to investigate how this secular model  compares with the results of the previously presented hydrodynamical simulations, we first consider the case with  $t_{dis}=10^8$ $T_{bin}$ so that the disc mass remains almost constant over the considered evolution times of
$~10^5$ $T_{bin}$, and we also assume $a_p=3.45$ initially such that the migration is almost stalled due to the action of the corotation torque.
Fig. \ref{fig:secular} shows the planet inclination $i_p$ as a function of the precession angle $\beta_p$ (the phase portrait) for initial disc masses ranging from $0.1$ to $20$ MMSN. The upper panel corresponds to the phase portrait that is 
obtained when the resonant term is not taken into account. For $M_D\le 2$ MMSN, the precession angle  $\beta_p$ circulates, while it librates for $M_D=20$ MMSN, resulting in a very small change in the planet inclination. This is consistent with the expectation that the planet precession becomes locked to the disc precession for massive discs (Fragner et al 2011; Lubow \& Martin 2016). Not surprisingly, the change in $i_p$ is also  modest for very  low-mass discs, and the precessional torque 
exerted by the binary makes $\beta_p$ circulate in that case with precession frequency given by Eq. \ref{eq:omegap}.

\begin{figure}
\centering
\includegraphics[width=\columnwidth]{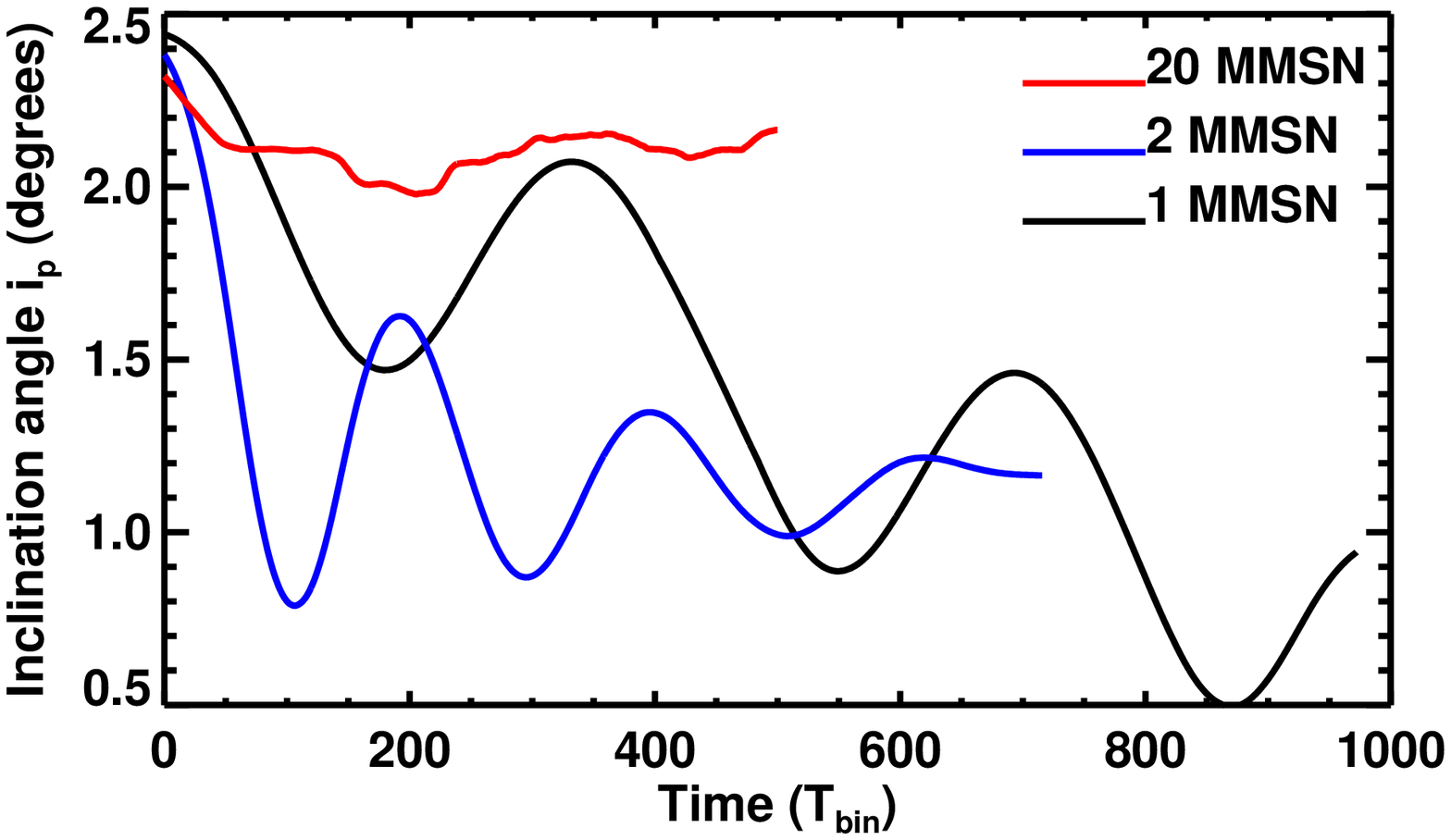}
\includegraphics[width=\columnwidth]{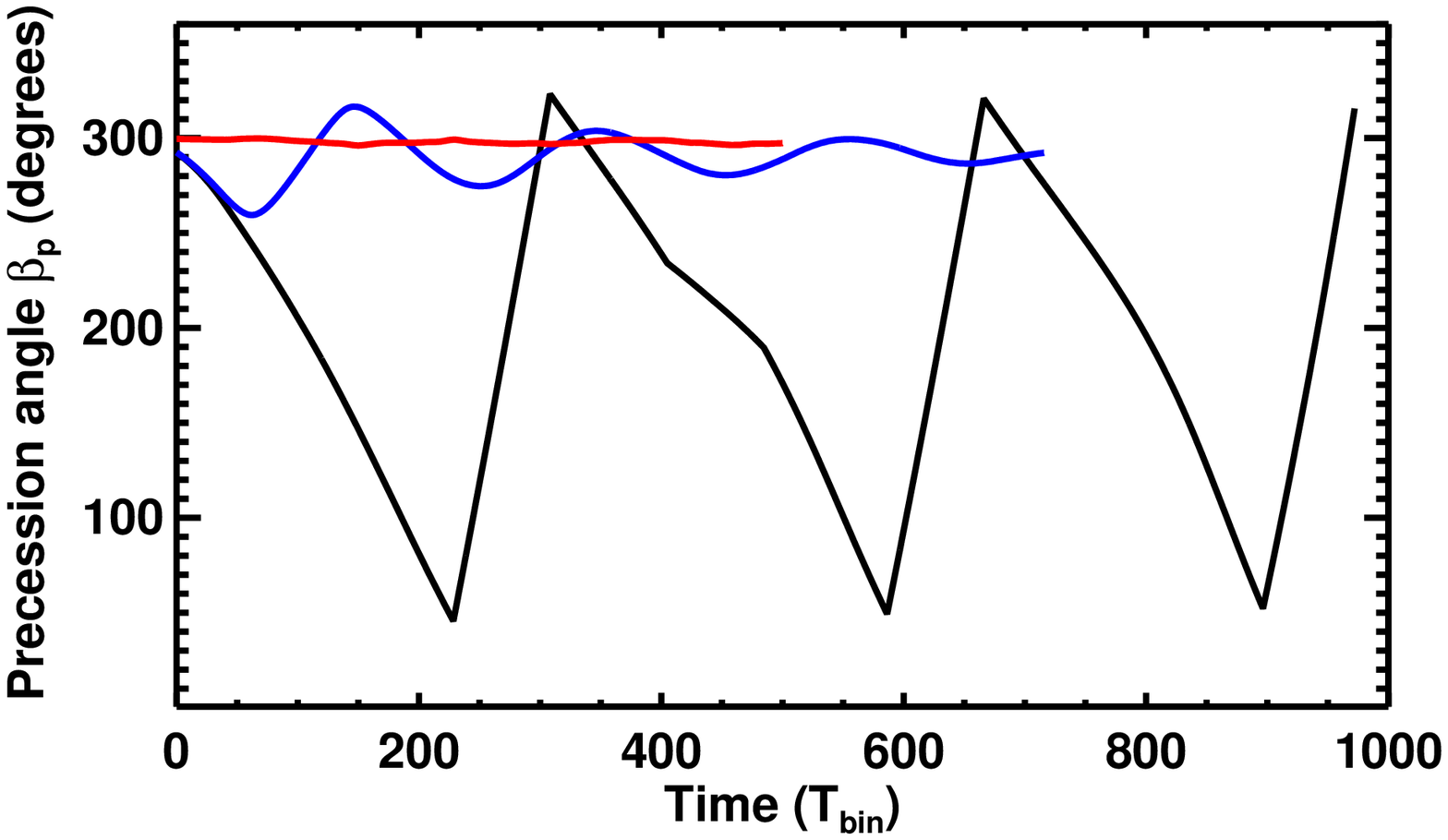}
\caption{ Time evolution of the planet inclination $i_p$ (upper panel) and planet precession angle $\beta_p$ (lower panel) for Model $1$ and for different disc masses.}
\label{fig:iversusm}
\end{figure}

The lower panel of Fig. \ref{fig:secular} corresponds to the case where resonant effects are considered. The oscillations 
for the tilt and precession angles are clearly damped in that case, consistently with what is observed in 
Fig. \ref{fig:iversusm}.  For $M_D\ge2$ MMSN, the final values for $i_p$ and $\beta_p$ are in very good agreement with the  results of hydrodynamical simulations. For $M_D=1$ MMSN, however, the hydrodynamical simulation shows circulation 
of the precession angle and continuous decrease in the planet inclination whereas in the  lower panel 
of Fig. \ref{fig:secular}, the precession angle is already librating due to damping effects after one synodic precession period. Nevertheless, the case corresponding to $M_D=0.1$ MMSN confirms that for sufficiently low mass discs, the resonant interaction with the disc causes the planet orbit to become almost perfectly aligned with respect to the binary orbit. 

\begin{figure}
\centering
\includegraphics[width=\columnwidth]{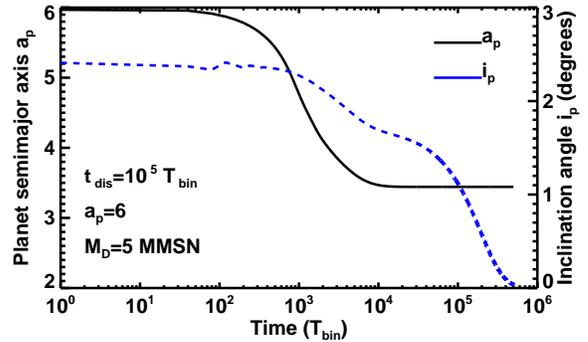}
\caption{Time evolution of the planet semimajor axis $a_p$ and inclination $i_p$ resulting from the secular model 
with $a_p=6$ initially and gas  dissipation timescale $t_{dis}=10^5$ $T_{bin}$.  }
\label{fig:secularm}
\end{figure}

The time evolution of the planet semi-major axis, $a_p$, and inclination, $i_p$, for a disc dissipation timescale of  $t_{dis}=10^5$ $T_{bin}$ and $a_p=6 \times a_{bin}$ initially is presented in Fig. \ref{fig:secularm} to illustrate how the migration of a planet in an inclined, dissipating disc around a central binary might evolve.  The idea here is not to fit any particular system, but to illustrate general effects.  As expected, coplanarity with the disc is  approximately maintained at early times, although there is a slight decrease in the relative planet-disc tilt due to the increase in the precessional torque exerted by the binary as the planet migrates inward. The planet is then observed to reach the edge of the truncated cavity at $t\sim 10^4$ $T_{bin}$ where it remains trapped, while the planet--binary relative inclination decreases as the disc disperses. At $t=10^5$ $T_{bin}$, the disc mass is $M_D\sim 1.8$ MMSN and $i_p\sim 1.2^\circ$, which is in good agreement with the results  of the hydrodynamical simulations (see Fig. \ref{fig:iversusm}). The subsequent evolution involves  continuous damping of the planet towards the binary plane as the disc mass decreases, which is again consistent with the results of both the hydrodynamical simulations and the secular model  with 
$t_{dis}=10^8$ $T_{bin}$ (see Fig. \ref{fig:secular}).

 \section{Discussion and conclusion}
We have presented the results of 3D hydrodynamical simulations that examine the migration and orbital evolution of planets orbiting around a system with stellar and planetary parameters intended to match the Kepler-413 circumbinary planet system.  We began our study by considering the evolution of circumbinary discs with large outer radii in the absence of embedded planets, assuming that the disc equatorial planes were initially moderately misaligned with respect to the orbital plane of the binary.  Our aim was to examine the effect of the tidal and
precessional torque exerted by the binary on the disc structure as a function of disc parameters. Most of our models have constant aspect ratio $h=0.05$, and we studied how the disc response varies as a function of the binary inclination $\gamma_{bin}$ and viscosity parameter $\alpha$. Binary inclinations of $\gamma_{bin}=2.5, 5, 8 ^\circ$  were studied and we considered values for the viscosity parameter of $\alpha=10^{-4}, 4\times 10^{-3}, 10^{-2}$. 

Our results confirm that in the wave-like regime for warp propagation, corresponding to $h>\alpha$, a small slowly-varying warp is set up in the disc due to bending waves communicating efficiently across the full radius of the disc.  The inner regions of the disc models are torqued towards alignment with the binary (with final tilt angles here being about half of the tilt at the outer edge of the disc). In that case, the disc achieves a quasi-steady state in the inner regions with $r \lesssim \;10 a_{bin}$, where $a_{bin}$ is the binary semi-major axis, after $\sim 1000$ binary orbital periods.  This quasi-equilibrium state is characterized by the disc maintaining a very small warp and twist, in addition to exhibiting almost no precession due to its large radial extent. 
The inner regions of the discs are tidally truncated, such that a low density inner cavity is formed, and are torqued towards alignment with the binary, with final tilt angles here being about half of the tilt at the outer edge of the disc. The disc structure is found to depend only weakly on the binary inclination. An equilibrium configuration is also achieved in the diffusive warp propagation regime with $\alpha> h$. In agreement with previous work (e.g. Larwood et al. 1996; Fragner \& Nelson 2010), we find that  in this regime the disc tends to be more twisted, simply because less efficient warp propagation requires a more distorted disc structure to be established before internal stresses counterbalance the effect of the binary precessional torque. 

Having obtained quasi-equilibrium (or at least slowly varying) disc configurations, we then examined the dynamical evolution of planets that are initially embedded in the disc on orbits that are locally coplanar with the disc midplane and close the the outer edge of the tidally truncated inner cavity. We adopted a planet-binary mass ratio equivalent to that of Kepler-413b. We first considered a circumbinary disc model with constant aspect ratio $h=0.05$, viscosity parameter $\alpha=4\times 10^{-3}$ and whose inclination with respect to the orbital plane of the binary is $\sim 2.5^\circ$ at the initial position of the planet (the outer edge of the disc was inclined by $5^\circ$). Consistently with previous work, we find stalling of migration once the eccentricity of the planet becomes high enough to induce a Lindblad torque reversal. For a disc mass $M_d$ equivalent to $1$ MMSN, we also find that coplanarity with the circumbinary disc is not maintained. The rapid nodal precession induced by the binary causes the planet orbit to precess with respect to the disc, and hence to periodically become highly inclined with respect to it as the relative precession angle reaches $\pi$. Damping of the relative motion due to dynamical friction (or equivalently due to inclination damping due to resonant disc-planet interactions) causes the planet orbit to settle towards the binary orbit plane.

For  higher disc masses with $M_d\ge 2$ MMSN, however, the disc gravity starts to become important and changes the evolution dramatically. For $M_d =20$ MMSN, the disc gravity is strong enough to force the planet to remain orbiting very close to the disc midplane. For $M_d =2$ MMSN, after a transient period during which the inclination and precession angles of the planetary orbit are observed to undergo damped oscillations, the planet settles into an orbit with a nodal precession angle very similar to that of the local disc annuli, and with an inclination angle that is smaller that the disc, but not fully aligned with the binary orbit plane. 

We also considered a flared disc model with a small aspect ratio $h=0.01$ at the initial position of the planet, for which the planet is expected to open a gap. This resulted in a similar mode of evolution for $i_p$ for a disc mass of 1 MMSN. Because of the gap formation by the planet,  the orbital evolution was different in that case but was in good agreement with previous work, showing continuous inward migration deep inside the inner cavity until the planet enters a MMR with the binary, or the amount of gas in the vicinity of the planet becomes too small to make the planet migrate further in. 
 
Previous work has demonstrated the plausibility of a formation scenario for the {\it Kepler} circumbinary planets involving core formation far away from the binary, followed by migration to the edge of the tidally truncated cavity. Gas accretion could have occurred during the migration or after the planet has reached the cavity edge. In the context of this scenario, our results indicate that it is likely that a core formed at large distance from the binary would maintain coplanarity with the outer disc due to the weak precessional torque exerted by the binary. This coplanarity, however, would ultimately be lost once the disc mass starts to significantly decrease during one of the following scenarios: i) during disc dissipation after the planet has reached the edge of the cavity; ii) in the course of migration if the core formed in a relatively low-mass disc. 

The main result of this paper is that when we consider the long term evolution of misaligned circumbinary discs with embedded planets, during which the disc mass decreases due to accretion onto the star and/or because a photoevaporative or magneto-centrifugal wind removes mass, the planet will tend to align with the binary orbit plane even if the disc-binary system remains misaligned up until the end of the disc life time. Hence, it is not necessary for the disc to align with the binary in order for the planet to lie within the binary orbit plane. The picture presented in this paper may therefore represent an alternative scenario to that presented in Foucart \& Lai (2013), who suggest that realignment of binaries and circumbinary discs should occur rapidly. If our results are correct, and other processes do not intercede to misalign circumbinary planets, then these planets should have orbits that are preferentially close to being aligned with the binaries, and far from being isotropically distributed. This has important implications for assessing the occurrence rates of circumbinary planets, since they have been estimated to be more common than planets around single stars in their inclination distribution is close to being isotropic (Armstrong et al. 2014). We have shown that for a single planet orbiting a pair of stars in a circumbinary disc, even if the disc inclinations are themselves isotropic (an unlikely scenario perhaps), the planetary orbits will not be. 

 Regarding the Kepler-413 and Kepler-453 systems, which contain planets with modest inclinations estimated to be $\sim 2.5^\circ$, it is possible that rapid gas disc removal may have prevented these systems from fully aligning with the central binary.  Other mechanisms, such as planet-planet scattering (Rasio \& Ford 1996; Chatterjee et al. 2008) or nodal precession due to interaction with a distant inclined companion might also be invoked to interpret the slight misalignments observed.  We also note that we have only considered
circumbinary discs that are slightly misaligned with respect to the binary orbit, and the case of higher disc-binary tilts needs to be investigated. For larger misalignment angles $i\gtrsim 20^\circ$, tilt oscillations between the planet and the disc can induce the inclination between the planet and the disc to increase above 
  $40^\circ$ because the planet precesses with respect to the disc, resulting in the formation a Kozai-Lidov planet (Martin et al. 2016). Considering larger binary inclinations will allow us to examine this and other effects in the contex of circumbinary planets, and will be the subjects of future work. 
\section*{Acknowledgments}
Computer time for this study was provided by the computing facilities MCIA (M\'esocentre de Calcul Intensif Aquitain) of the Universite de Bordeaux and by HPC resources of Cines under the allocation A0010406957 made by GENCI (Grand Equipement National de Calcul Intensif).  

\end{document}